\documentclass{report}
\usepackage[utf8]{inputenc}
\usepackage{mathtools}
\usepackage{caption}
\usepackage{subcaption}
\usepackage{amsfonts}
\usepackage{amsmath}
\usepackage{geometry}
\usepackage{tabularx}
\usepackage{multicol}
\usepackage{comment}
\usepackage{titlepic}
\usepackage[export]{adjustbox}
\newcommand{\beq}{\begin{equation}}
\newcommand{\eeq}{\end{equation}}

\begin{document}

\begin{titlepage}

\includegraphics[width=0.3\textwidth, left]{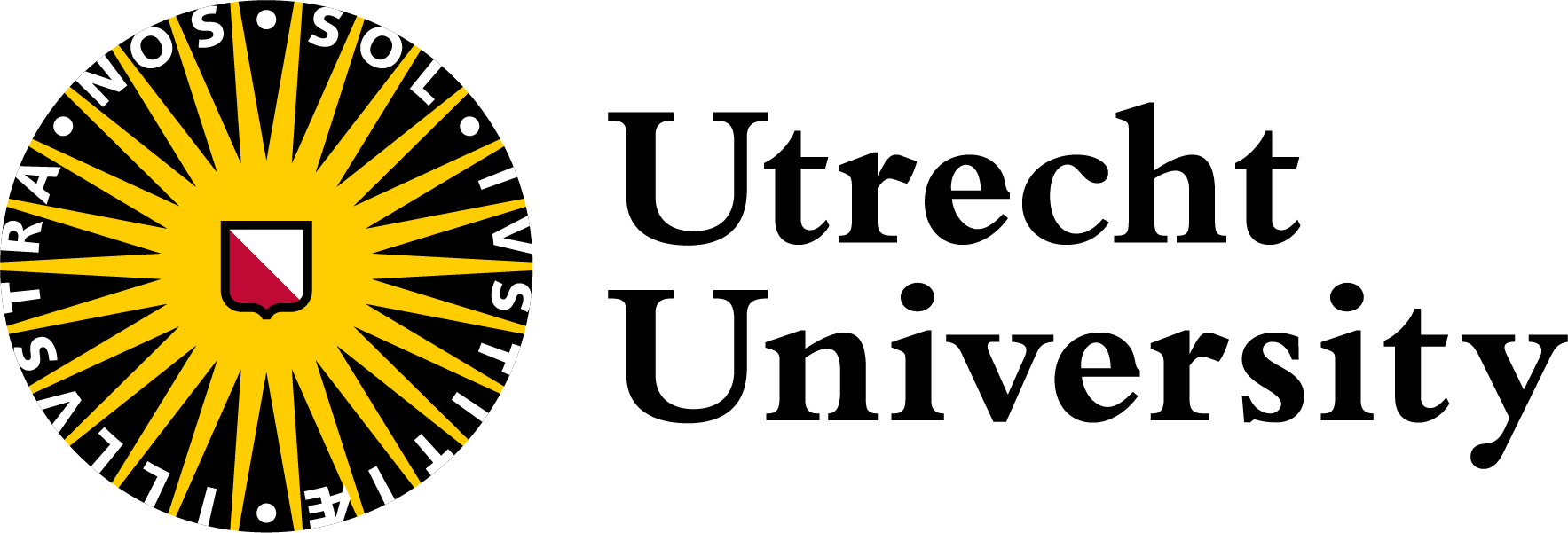}

\begin{centering}
    \vspace{1.5cm}
    
    {\huge Modeling a Financial System with Memory via Fractional Calculus and Fractional Brownian Motion}
    
    \vspace{1.5cm}
    
    Patrick Geraghty\\[0.5cm]{}\small Advisors: Prof. Dr. Cristiane Morais Smith, Prof. Dr. Cornelis Oosterlee, Robin Verstraten MSc\\[0.5cm]{}\small Institute of Theoretical Physics and Mathematical Institute
    
    \vspace{0.5cm}
    
    July 10th, 2022
    
    \begin{figure}[b!]
        \centering
        \includegraphics[width=0.8\textwidth]{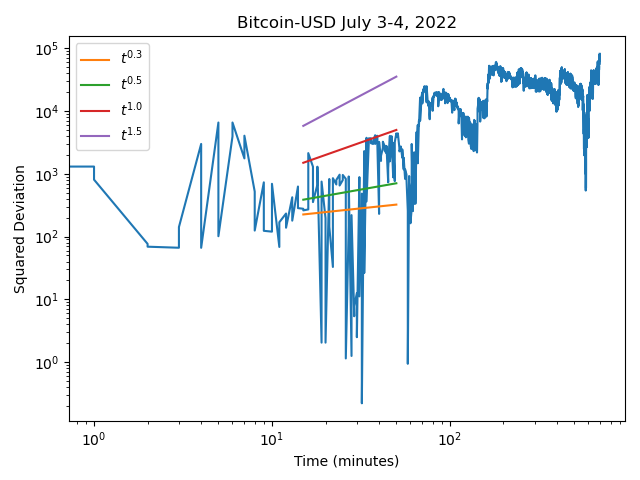}
    \end{figure}
    
\end{centering}

\end{titlepage}

\begin{abstract}
Financial markets have long since been modeled using stochastic methods such as Brownian motion, and more recently, rough volatility models have been built using fractional Brownian motion.  This fractional aspect brings memory into the system.  In this project, we describe and analyze a financial model based on the fractional Langevin equation with colored noise generated by fractional Brownian motion.  Physics-based methods of analysis are used to examine the phase behavior and dispersion relations of the system upon varying input parameters.  A type of anomalous marginal glass phase is potentially seen in some regions, which motivates further exploration of this model and expanded use of phase behavior and dispersion relation methods to analyze financial models.
\end{abstract}

\section*{Acknowledgements}
I would like to thank my supervisors Prof. Dr. Cristiane Morais Smith and Prof Dr. Cornelis Oosterlee for their support.  I would also like to thank Robin Verstraten for always being ready to answer questions and talk through ideas.  A research project that branches between two very different subjects is bound to be difficult, but I am very thankful for all the enthusiasm and effort put into understanding each other and finding the similarities in these two methodologies.  This project started off as a random idea in my head and it was through the contributions of everyone involved that it became a reality, so thank you for working with me all this way.

I would also like to thank Dr. Tanner Crowder, who back in the summer of 2018, gave me the opportunity, as an undergraduate research intern, to pick what topics to study for the summer.  It was there that I first encountered fractional calculus and realized how incredible the subject is.  From that first experience, I got the drive to start this research project into financial applications of fractional calculus.

This project also would not have been possible without the invaluable support of my friends, in the physics program, at the Vault, and everywhere else.  Thanks especially to Leonardo H\"ugens, we did nearly this entire program together helping each other through the worst of the pandemic and the best of times.

\tableofcontents

\chapter{Introduction}
While many may not think of financial modeling as a poetic field, in 60 BCE, Lucretius published a scientific poem \cite{lucretius-poem} about particles being "moved by blows that remain invisible," a phenomenon which 2000 years later would be used to describe financial systems.  Even before Einstein published his seminal paper \cite{einstein} on the atomic theory, Louis Bachelier used Brownian motion for a stochastic analysis of the stock market \cite{bachelier}.  It was this work that many consider to be the birth of mathematical finance, and since then Brownian motion has been the core of the majority of financial models, \cite{black-scholes}-\cite{brownian-markets}.

In the first half of the 20th century, Harold Hurst was using statistics to study the varying levels of the Nile river and found that the statistics scaled with respect to the number of samples to some power.  These statistics could be used to describe all kinds of systems including walks.  For a random walk, this power was $1/2$; however, Hurst found powers ranging up to one \cite{og-hurst}.  Over the next couple of decades, more and more study went into these scaling relations by a variety of mathematicians \cite{fbm-history}.  The whole field came together in 1968, when Benoit Mandelbrot and John Van Ness published their seminal work \cite{mandelbrot-fbm} giving the name Fractional Brownian Motion to walks that scale with powers other than $1/2$.  The scsaling power was then named the Hurst parameter, $H$.  For $H=1/2$, regular Brownian motion is reproduced, as each step is random and hence independent of all previous ones.  However, for values of $1/2<H<1$ and $0<H<1/2$, the next step becomes positively or negatively correlated with the past, respectively.  These correlations lead to paths that have memory of their past that impacts their future.  This method, however, is not the only way to define Brownian motion.

Early in the 1900s, Paul Langevin developed a differential equation that describes the evolution of a system under the impact of a random force \cite{og-langevin}.  The equation is effectively Newton's second law where the forces acting on the trace particle are the random force and a friction force.  When the system is in equilibrium, these forces are related to each other via the fluctuation-dissipation theorem which will be discussed in detail later on.  When the random force in this equation has a Gaussian probability distribution, the equation exactly describes the Brownian motion of a small particle moving in a fluid.  Then how can fractional Brownian motion emerge?

When calculus was initially being developed simultaneously by Newton and Leibniz, there was a letter sent from Leibniz to l'H\^{o}pital in 1695 \cite{hilfer} talking of a potential form of a fractional derivative: "this is an apparent paradox from which, one day, useful consequences will be drawn."  Through the next centuries, work on developing fractional calculus slowly continued \cite{frac-calc-hist}, up to 1969 when Michele Caputo developed a definition that sparked new life into the field \cite{caputo-deriv}.  This definition included an integral over all previous times, thus showing a connection to memory in the system.  Before this, there were many definitions of fractional derivatives which were not equivalent.  By taking the Langevin differential equation and extending the friction term to be a Caputo fractional derivative of order $\alpha$, Brownian motion with memory can be modeled: fractional Brownian motion.

The idea of using fractional Brownian motion to model financial markets is not new, models for stochastic volatility were first described in the late 1980s and early 1990s by Hull and White \cite{hull-white}, Heston \cite{heston}, and others \cite{rough-bergomi}-\cite{malliavin-book}.  This allowed for an extension of the deterministic volatility model famously written by Black and Scholes \cite{black-scholes}.  With the extension, volatility surfaces could be modeled that were more representative of real market data especially in the era of high-frequency trading \cite{volat-rough}.

All of these models begin with financial axioms and the analysis is all based in finance.  In this project, we chose our starting point to be the financial model described by Rama Cont and Jean-Philippe Bouchaud \cite{fin-lang}, then we used physics-based methods to extend the model to fractional Brownian motion such that it includes memory.  These physics methods allow us to analyze the model by examining the phase behavior and dispersion relations of the system as a function of the input financial parameters.

This report begins with an introduction to fractional calculus in Chapter \ref{chap:frac-calc}, followed by a description of the equivalent methods of finance-based stochastics and physics-based differential equations to model fractional Brownian motion in Chapter \ref{chap:fbm}.  In Chapter \ref{chap:fin-model}, we explain the financial model built by Cont and Bouchaud and extend it to fractional order.  Then, in Chapter \ref{chap:num-sim}, we examine how to numerically simulate fractional Brownian motion and how to solve fractional differential equations using a Monte Carlo method.  Finally, in Chapter \ref{chap:analysis}, we analyze the full financial model of a market with memory.

\chapter{Fractional Calculus}
\label{chap:frac-calc}
There is no single method of extending regular calculus to fractional orders, which means that there are a variety of definitions of fractional derivatives that are not always equivalent.  This project focuses on three of the most common versions: the Riemann-Liouville form, the Gr\"uwald-Letnikov form, and the Caputo form by using Refs. \cite{podlubny} and \cite{hilfer}.  Other useful sources for fractional calculus include Refs. \cite{frac-calc-1}-\cite{frac-calc-7}.

\section{Useful Functions}
In order to introduce these forms, we first define a few useful mathematical functions.
\subsection*{Gamma Function}
One of the most prevalent functions in fractional calculus is the Gamma function which is an extension of the factorial function defined by:

\beq
\Gamma(z) = \int_{0}^{\infty} \textrm{dt} e^{-t} t^{z-1}, \quad Re(z)>0,
\label{eq:gamma-func-def}
\eeq
where $z$ can be non-integer and even imaginary.  If one looks at $\Gamma(1)=1$ and uses that $\Gamma(z+1)=z\Gamma(z)$, it can be seen that for $n \in \mathbb{Z}$, one finds $n$!$=\Gamma(n+1)$.

\begin{figure}
     \centering
     \begin{subfigure}[b]{0.45\textwidth}
         \centering
         \includegraphics[width=\textwidth]{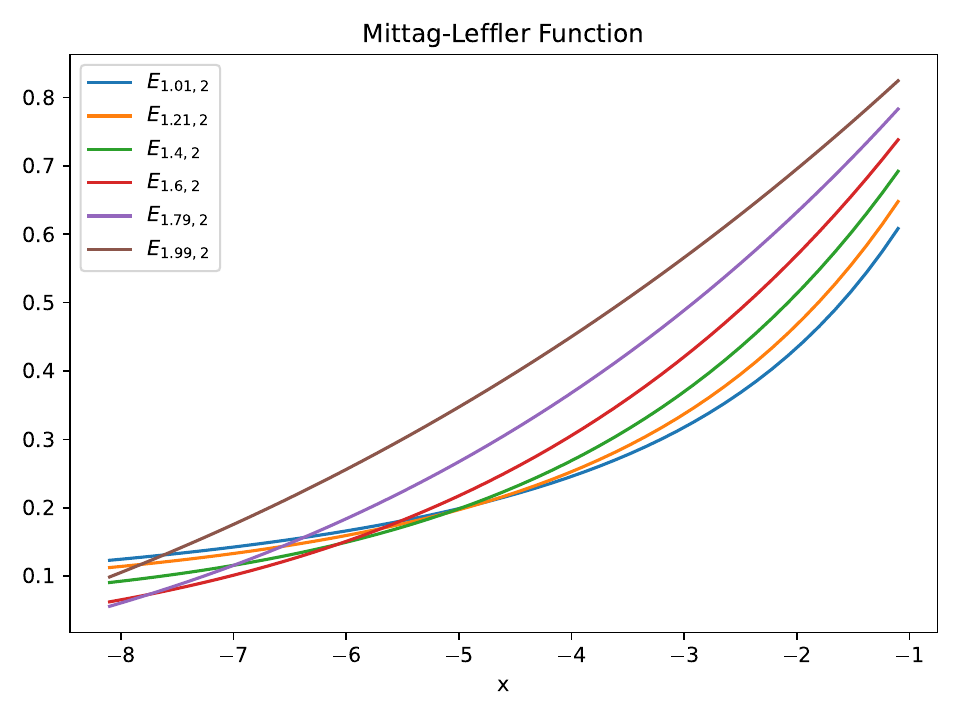}
         \subcaption{}
     \end{subfigure}
     \hfill
     \begin{subfigure}[b]{0.45\textwidth}
         \centering
         \includegraphics[width=\textwidth]{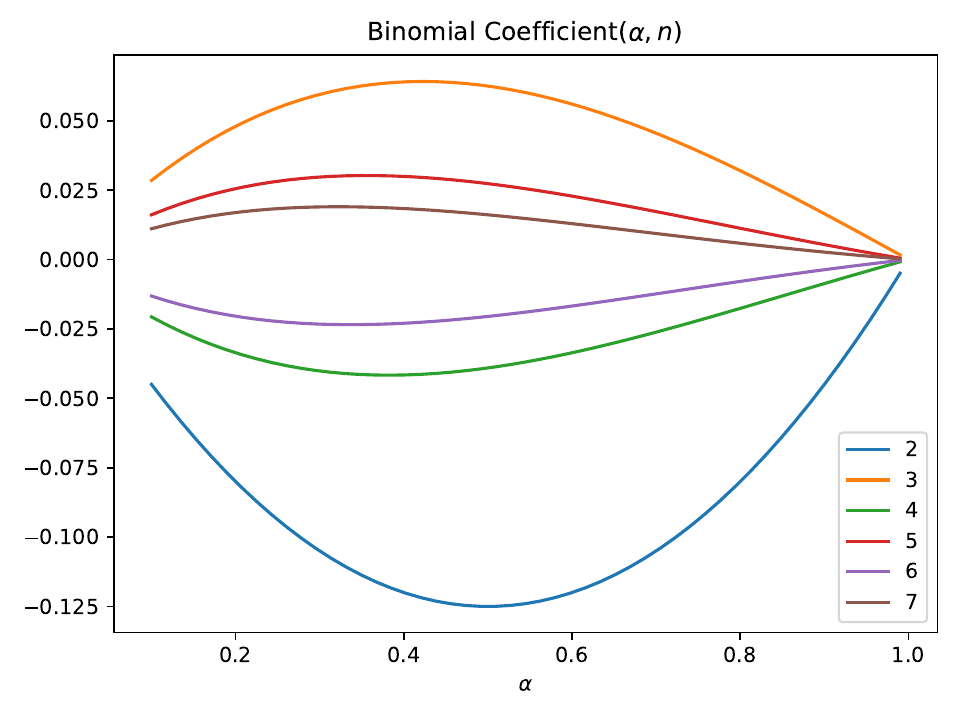}
         \subcaption{}
     \end{subfigure}
        \caption{Figure (a) shows the Mittag-Leffler function from Equation \ref{eq:mitleff}, for a variety of parameters that will become useful.  Figure (b) shows the Binomial Coefficient, which has a switching behavior depending on whether $n$ is even or odd.}
        \label{fig:useful-funcs}
\end{figure}

\subsection*{Mittag-Leffler Function}
The next useful function, which appears nearly as often in fractional calculus, is the two-parameter Mittag-Leffler function.  This function is one way to extend the exponential function.

\beq
E_{\alpha,\beta} (z) = \sum_{k=0}^{\infty} \frac{z^k}{\Gamma(\alpha k + \beta)}.
\label{eq:mitleff}
\eeq
It can be seen that $E_{1,1}(z)$ = $e^z$ and $E_{a,b} (0)$ = $1/ \Gamma(b)$.  Figure \ref{fig:useful-funcs}(a) shows $E_{\alpha,2} (z)$ for a range of $\alpha$ and $z$ values.  The plot shows $\beta$ = 2 because this is the version of the Mittag-Leffler function that will be most useful for our purposes.  In this case, $z>0$ because the Gamma function discussed previously is only defined for positive values.

\subsection*{Binomial Coefficient}
By using the above definition of the Gamma function, the binomial coefficient can be extended:

\beq
\begin{pmatrix}
\alpha \\ n
\end{pmatrix}
= \frac{\Gamma(\alpha + 1)}{\Gamma(n+1)\Gamma(\alpha - n + 1)},
\label{eq:binom-coeff-def}
\eeq
where $\alpha \in \mathbb{R}$ and in the cases to follow $n \in \mathbb{Z}$.  Figure \ref{fig:useful-funcs}(b) shows the binomial coefficient for a range of $\alpha$ and $n$ values.  It is worth noticing that the function is either negative or positive depending on whether $n$ is even or odd.

\section{Riemann-Liouville Form}
In order to define the fractional derivative, we start by defining the fractional integral.  This is done by iterating a first-order integral $n$ times:

\beq
I_{a+}^{n} f(t) = \int_{a}^{t} \int_{a}^{t_{1}} \cdots \int_{a}^{t_{n-1}} f(t_n) dt_n \ldots dt_2 dt_1 = \frac{1}{\Gamma(n)} \int_{a}^{t} d\tau (t-\tau)^{n-1} f(\tau),
\eeq
where $n$ is an integer and the $a+$ on the left-hand side means that $a$ is the lower bound.  A derivative is the inverse of an integral as seen by

\beq
\frac{d}{dx}\int dx f(x) = f(x) \rightarrow D \big[ I[f(x)] \big] = f(x) = D^{1} \big[ D^{-1} [f(x)] \big] \rightarrow I = D^{-1}.
\eeq
Hence, in this operator form, the iterated integral $I_{a+}^{n} = D_{a+}^{-n}$.  By extending $n \rightarrow \alpha$, where $\alpha \in \mathbb{R}$, the \textit{Riemann-Liouville fractional integral} of order $\alpha$ reads:

\beq
{}^{RL} D_{a+}^{-\alpha} f(t) = \frac{1}{\Gamma(\alpha)} \int_{a}^{t} d\tau (t-\tau)^{\alpha-1} f(\tau).
\eeq
Using the operator language from before, the \textit{Riemann-Liouville fractional derivative} reads: $D^{n} D^{\alpha-n} = D^{\alpha}$, where $n-1 < \alpha < n$,

\beq
{}^{RL} D_{a+}^{\alpha} f(t) = \frac{d^n}{dt^n} {}^{RL} D_{a+}^{\alpha - n} f(t) = \frac{1}{\Gamma(n-\alpha)} \frac{d^n}{dt^n} \int_{a}^{t} d\tau (t-\tau)^{n-\alpha-1} f(\tau).
\label{eq:rl-deriv}
\eeq
Let us consider an example function $f(t)=(t-a)^{\sigma}$:

\begin{align*}
{}^{RL} D_{a+}^{\alpha} f(t) &= \frac{1}{\Gamma(n-\alpha)} \frac{d^n}{dt^n} \int_{a}^{t} d\tau (t-\tau)^{n-\alpha-1} (\tau-a)^{\sigma}, \qquad z = \frac{\tau-a}{t-a} \quad \rightarrow \quad dz = \frac{d\tau}{t-a}, \\
&= \frac{1}{\Gamma(n-\alpha)} \frac{d^n}{dt^n} \bigg[ (t-a)^{n-\alpha+\sigma} \int_{0}^{1} dz (1-z)^{n-\alpha-1}z^{\sigma} \bigg], \\
&= \frac{1}{\Gamma(n-\alpha)} \frac{d^n}{dt^n} \bigg[ (t-a)^{n-\alpha+\sigma} \frac{\Gamma(\sigma+1)\Gamma(n-\alpha)}{\Gamma(n-\alpha+\sigma+1)} \bigg], \\
&= \frac{\Gamma(\sigma+1)}{\Gamma(n-\alpha+\sigma+1)} \frac{d^n}{dt^n} \bigg[ (t-a)^{n-\alpha+\sigma} \bigg], \\
&= \frac{\Gamma(\sigma+1)}{\Gamma(n-\alpha+\sigma+1)} \frac{\Gamma(n-\alpha+\sigma+1)}{\Gamma(\sigma-\alpha+1)}  (t-a)^{\sigma-\alpha}, \quad \textrm{and} \\
&= \frac{\Gamma(\sigma+1)}{\Gamma(\sigma-\alpha+1)}  (t-a)^{\sigma-\alpha}.
\end{align*}
The result is relatively similar to the integer order case where the order of the polynomial decreases by the order of the derivative.  The coefficient is the main difference which is typically made of a factorial function in the integer case.  With the fractional derivative, the coefficient is now the non-integer extension of the factorial: the gamma function.

\begin{figure}
     \centering
     \begin{subfigure}[b]{0.45\textwidth}
         \centering
         \includegraphics[width=\textwidth]{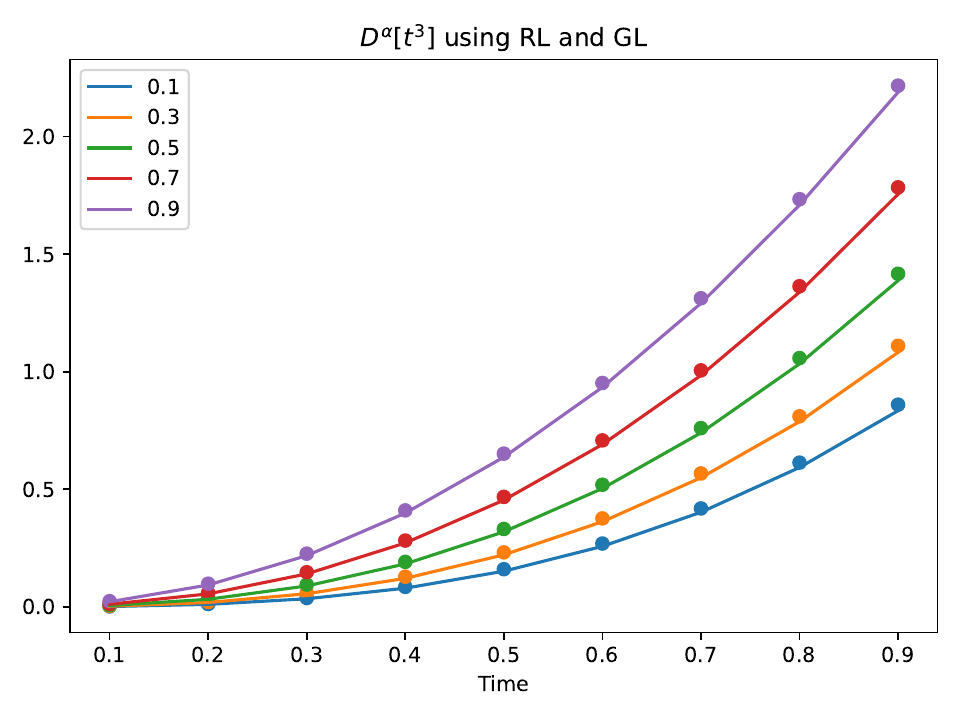}
         \subcaption{}
     \end{subfigure}
     \hfill
     \begin{subfigure}[b]{0.45\textwidth}
         \centering
         \includegraphics[width=\textwidth]{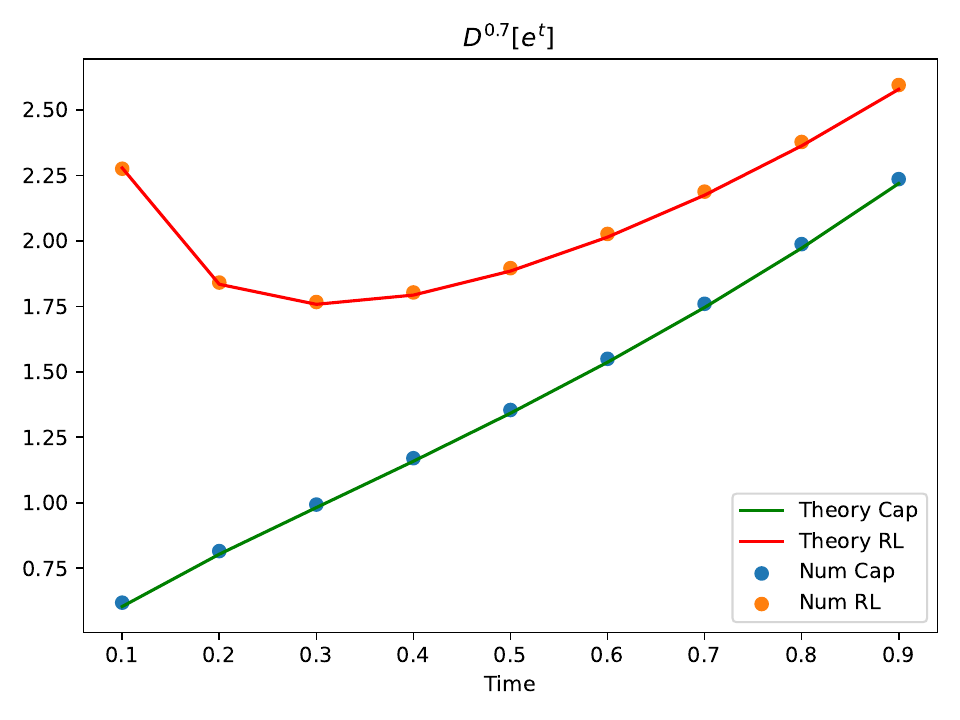}
         \subcaption{}
     \end{subfigure}
        \caption{Figure (a) shows the analytical (Riemann-Liouville) and numerical (Gr\"unwald-Letnikov) fractional derivative of $t^3$ for a variety of derivative orders using Equations \ref{eq:rl-deriv} and \ref{eq:gru-let}.  Figure (b) shows, numerically and analytically, how Equation \ref{eq:trans-rl-2-cap} can be used to calculate the Caputo derivative from the Riemann-Liouville form.}
        \label{fig:frac-derivs-num}
\end{figure}

\section{Gr\"unwald-Letnikov Form}
The next version of the fractional derivative to be introduced starts from the limit definition of the derivative and in the end is equivalent to the Riemann-Liouville form.  The limit definition states that

\beq
\frac{df}{dt} = \lim_{h\rightarrow0} \frac{f(t)- f(t-h)}{h},
\eeq
and by iterating this expression:

\beq
\frac{d^2f}{dt^2} = \lim_{h\rightarrow0} \frac{f'(t)- f'(t-h)}{h} = \lim_{h\rightarrow0} \frac{f(t)- 2f(t-h)+f(t-2h)}{h^2}.
\eeq
It can be seen by induction that the iterated derivative can be generalized to:

\beq
\frac{d^nf}{dt^n} = \lim_{h\rightarrow0} \frac{1}{h^n} \sum_{j=0}^{n} (-1)^j \begin{pmatrix} n \\ j \end{pmatrix} f(t-jh).
\eeq
Generalizing this to fractional derivatives, we send $n \rightarrow \alpha$ whenever possible:

\beq
{}^{GL} D_{a+}^{\alpha} = \lim_{h\rightarrow0} \frac{1}{h^{\alpha}} \sum_{j=0}^{n} (-1)^j \begin{pmatrix} \alpha \\ j \end{pmatrix} f(t-jh),
\eeq
where $\alpha<n$ and the upper limit of the sum is still $n$ because one cannot sum over fractions.  However, there is still ambiguity in what $n$ is doing.  Thus, let us consider the fractional integral by sending $\alpha \rightarrow -\alpha$.  This is done by knowing that $\begin{pmatrix} -p \\ j \end{pmatrix}=(-1)^j \begin{pmatrix} p \\ j \end{pmatrix}$ where $j$ is an integer.

\beq
f^{(-\alpha)}(t) = \lim_{h\rightarrow0} h^{\alpha} \sum_{j=0}^{n} \begin{pmatrix} \alpha \\ j \end{pmatrix} f(t-jh).
\eeq
When the limit is taken in Equation 2.12, the result is trivial. However, if the limit $n\rightarrow\infty$ is taken as well, by using the notation of Riemann integration a non-trivial result is found.  Taking the total width, $nh$, to be $t-a$, the resulting expression is a Riemann integral of order $\alpha$, i.e.,

\beq
f^{(-\alpha)}(t) = \lim_{\substack{h\rightarrow0 \\ nh \rightarrow t-a}} h^{\alpha} \sum_{j=0}^{n} \begin{pmatrix} \alpha \\ j \end{pmatrix} f(t-jh) = {}^{GL} D_{a+}^{-\alpha} f(t).
\label{eq:gru-let}
\eeq
The definition of this form started with the limit definition of the derivative and when negative orders were taken, the discrete Riemann integral was found.  By looking at the limiting case as $h \rightarrow 0$, it is known that the discretized Riemann integral is equivalent to the analytical form.  In the same way, the discrete, fractional, Gr\"unwald-Letnikov form is equivalent to the Riemann-Liouville form of fractional derivatives and integrals.

The question then becomes why do we need both forms if they are equivalent?  The answer is that the Riemann-Liouville form is useful for analytical work and the Gr\"unwald-Letnikov form is useful for numerical work.  The power law term in the integral of the Riemann-Liouville form diverges as it approaches the upper bound and so, numerically, it is impossible to work with.  Thus using the discretized version, the Gr\"unwald-Letnikov form, it is possible to numerically approximate the fractional derivative or integral of a function.  The left plot in Figure \ref{fig:frac-derivs-num} shows the analytical form of the fractional derivative of $y=t^3$ by the Riemann-Liouville method and the numerical approximation using the Gr\"unwald-Letnikov method.

\section{Caputo Form}
The Caputo form is complementary to the the Riemann-Liouville form.  The key difference is that the sequence of the integral and the $\textrm{n}^{\textrm{th}}$-order derivative is swapped.

\beq
{}^{C} D_{a+}^{\alpha} f(t) = \frac{1}{\Gamma(n-\alpha)} \int_{a}^{t} d\tau \frac{f^{(n)}(\tau)}{(t-\tau)^{\alpha-n+1}}, \quad n-1 < \alpha < n.
\label{eq:caputo-deriv}
\eeq
The Caputo derivative can be compared to the Riemann-Liouville form by:

\beq
{}^{RL} D_{a}^{\alpha} f(t) = \sum_{k=0}^{n-1} \frac{f^{(k)}(a) (t-a)^{k-\alpha}}{\Gamma(k-\alpha+1)} + {}^{C} D_{a}^{\alpha} f(t).
\label{eq:trans-rl-2-cap}
\eeq
In order to obtain this relationship, the integer-order derivative on the outside of the Riemann-Liouville form needs to be moved past the power law term and into the integral.  This is done via repeated integration by parts.  The Riemann-Liouville and Caputo derivatives of the exponential function, $e^t$, are shown both analytically and numerically in Figure \ref{fig:frac-derivs-num}(b).  The numerical Caputo derivative is found by using first the Gr\"unwald-Letnikov form and then Equation \ref{eq:trans-rl-2-cap}.

\chapter{Fractional Brownian Motion}
\label{chap:fbm}
Fundamentally, Brownian motion is understood as the stochastic interactions between a trace particle and a bath of other particles \cite{caldeira-leggett}.  The trace particle's interaction with each particle in the bath causes the trace particle to move a small amount.  As the trace particle has more and more interactions, a path is created.  It is this specific type of path that is called Brownian motion \cite{bm}.

When the trace particle interacts with a bath particle, the trace particle moves by some small increment.  This increment is only dependent on the most recent interaction the trace particle has, so what happens if this is changed to include other past interactions?  Then, a system could be build where the movement of the trace particle is dependent on how it has moved previously: memory.  This inclusion of memory allows us to generalize Brownian motion to fractional Brownian motion.

As discussed in the Introduction, Brownian motion has been connected to both physics, Einstein \cite{einstein}, and finance, Bachelier \cite{bachelier}, since its initial mathematical description, but the two fields have diverged in the formalism they each use.  We will begin with the physics approach which is based on differential equations and uses the parameter, $\alpha$, to describe fractional Brownian motion.  Then, we will describe finance's approach where fractional Brownian motion is examined as a stochastic process with the fractional Hurst parameter, $H$.
\section{Langevin Equation Approach}\label{sec:lang-approach}
In order to describe the physics approach to fractional Brownian motion, we follow Refs.\cite{robin-time-glass} and \cite{cl-1}-\cite{cl-7} in their derivation of the Langevin equation.  This derivation is done by starting with the Caldeira-Leggett model of quantum dissipation \cite{caldeira-leggett}.  The derivation starts by building a Lagrangian for the model which comes in four parts: the trace particle of our system, $\mathcal{L_S}$, the bath, $\mathcal{L_B}$, the interaction, $\mathcal{L_I}$, and a counter-term, $\mathcal{L_{CT}}$.

\beq
\mathcal{L} = \mathcal{L_S} + \mathcal{L_B} + \mathcal{L_I} + \mathcal{L_{CT}},
\eeq
\beq
\mathcal{L_S} = \frac{1}{2}M \dot{x}^2,
\eeq
\beq
\mathcal{L_B} = \sum_j \frac{1}{2} m_j \dot{x_j}^2 - \frac{1}{2} m_j \omega_{j}^{2} x_{j}^{2},
\eeq
\beq
\mathcal{L_I} = \sum_j C_j x_j x, \quad \textrm{and}
\eeq
\beq
\mathcal{L_{CT}} = -\sum_j \frac{1}{2} \frac{C_{j}^{2}}{m_j \omega_{j}^{2}} x^2,
\eeq
where the trace particle has mass $M$ and the bath is composed of non-interacting harmonic oscillators with coordinates $x_j$, masses $m_j$, and natural frequencies $\omega_j$.  The coupling strength between the trace particle and a given oscillator from the bath is $C_j$.  Already, the model looks similar to a regular Brownian motion system where the bath is represented by harmonic oscillators.  The counter-term part of the Lagrangian is included to keep the potential well-behaved under renormalization.  This would be an unphysical phenomenon and needs to be avoided.

Going through the Euler-Lagrange calculations inspired by Refs \cite{cl-1}-\cite{cl-7}, the following differential equation is found:

\beq
M\ddot{x} + F_{fr} = f(t), \quad \textrm{where}
\label{eq:early-lang}
\eeq
\beq
F_{fr} = \sum_j \frac{d}{dt} \bigg\{ \frac{C_{j}^{2}}{m_j \omega_{j}^{2}} x(t) * \textrm{cos}(\omega_j t) \bigg\}, \qquad f(t) = \sum_j C_j \bigg[ \frac{\dot{x}_j(0)}{\omega_j} \textrm{sin}(\omega_j t) + q_j (0) \textrm{cos}(\omega_j t) \bigg],
\label{eq:early-fric-noise}
\eeq
where the operation $(*)$ is the convolution operator.  In Equation \ref{eq:early-lang}, $F_{fr}$ is the force due to friction and $f(t)$ is an opposing force entirely dependent on the bath components.  The Caldeira-Leggett model was proposed as a model of quantum dissipation.  For our purposes, we look only at the macroscopic impacts and not the quantum phenomena.  Since we have a Lagrangian system, energy is conserved.  In this way, the friction force between the trace particle and the bath removes energy from the trace particle and \textit{dissipates} that energy to the microscopic harmonic oscillators in the bath.  The force $f(t)$ is what dissipates the energy from the trace particle to the bath which is why it has opposite sign to the friction force.  The interaction here comes from the fluctuation-dissipation theorem which is a result of the equipartion of energies in the system \cite{equip-flucdiss}.

The conservation of energy along with equipartion implies that there should be some relationship between $F_{fr}$ and $f(t)$ which is not readily obvious in Equation \ref{eq:early-fric-noise}.  The key is to rewrite both of these expressions in terms of a spectral function $J(\omega)$:

\beq
J(\omega) = \textrm{Im} \mathcal{F} \bigg[ -i\theta(t-t') \bigg\langle \big[ \sum_j C_j x_j (t), \sum_{j'} C_{j'} x_{j'} (t) \big] \bigg\rangle \bigg],
\label{eq:spec-func}
\eeq
where $\mathcal{F}$ is the Fourier transform, Im means taking the imaginary component, $\theta$ is the Heaviside function, and $[*,*]$ is the commutator.  Using Equation \ref{eq:spec-func}, the expressions for $F_{fr}$ and $f(t)$ can be rewritten as:

\beq
F_{fr} = \frac{d}{dt} \bigg\{ \frac{2}{\pi} \int_{0}^{t} d\tau \int_{0}^{\infty} d\omega \frac{J(\omega)}{\omega} \textrm{cos} \big( \omega [t-\tau] \big) x(\tau) \bigg\},
\label{eq:fric-spect}
\eeq
\beq
\big\langle f(t) f(t') \big\rangle = \frac{2 k_B T}{\pi} \int_{0}^{\infty} d\omega \frac{J(\omega)}{\omega} \textrm{cos} \big( \omega [t-t'] \big),
\eeq
where $\langle f(t) f(t') \rangle$ is the time correlation function of $f(t)$ and $k_B T$ is the thermal energy of the system.  The thermal energy of the system is explained in more detail in Section \ref{sec:thermal-energy}.

The question now is what exactly is the form of the spectral function $J(\omega)$?  In Ref \cite{cl-1}, it is taken to be linear in $\omega$, at which point $F_{fr}$ behaves exactly as a typical friction force and the time correlation of $f(t)$ becomes that of a white-noise force.  The result is the regular Langevin equation

\beq
M \ddot{x} + \gamma \dot{x} = f(t),
\label{eq:lang-two}
\eeq
which desribes regular Brownian motion.  However, we want fractional Brownian motion, so instead we take $J(\omega)=A\omega^{\alpha}$, where $\alpha \in \mathbb{N}$.  As discussed in the Introduction, we essentially want to extend the integer-order time derivative in the friction term in Equation \ref{eq:lang-two} to fractional order, so we want a value for $A$ in $J(\omega)$ such that Equation \ref{eq:fric-spect} can be written as a Caputo fractional derivative of the form given in Equation \ref{eq:caputo-deriv}.  The expression for $J(\omega)$ that does just that is given by

\beq
J(\omega) = \gamma \frac{\pi}{2} \frac{\sec(\pi \alpha / 2)}{\Gamma(\alpha) \Gamma(1-\alpha)} \omega^{\alpha},
\label{eq:spectral-fract}
\eeq
where $\gamma$ is a constant that will be examined later on.  With this expression, we now have that $F_{fr}=\gamma {}^{C}D_{0}^{\alpha} x(t)$.  For $\alpha$ = 1, we recover the Langevin equation for regular Brownian motion which is exactly what is expected for a spectral function that is simply linear in $\omega$.

There is of course the question of how memory appears in the system and the answer comes from the fractional derivative.  The regular derivative is only a function of the current time and so there is no memory.  However, the fractional derivative is an integral over all of history up to the current time.  It is this integral formulation that provides the inclusion of memory.

Now that we have an expression for $J(\omega)$, its time correlation can be evaluated to see what kind of noise it generates.  From Ref. \cite{lutz}, it is known that the time covariance of coloured noise will scale as $t^{-\alpha}$, where $\alpha$ is the fractional derivative order,

\beq
\big\langle f(t) f(t') \big\rangle = \frac{\gamma k_B T}{\Gamma(1-\alpha)} \vert t-t' \vert^{-\alpha}. 
\label{eq:noise-covar-lang}
\eeq
Thus, the noise generated by a power law spectral function $J(\omega)$ will be coloured noise.  For $\alpha$ = 1, we can look at the limit and see that what returns is a Dirac delta of $(t-t')$ which is the expected time correlation for white noise.  All this together leads to the fractional Langevin equation:

\beq
M \frac{d^2x}{dt^2} + \gamma {}^{C}D_{0}^{\alpha} \big[ x(t) \big] = f(t),
\label{eq:lang-3}
\eeq
which describes fractional Brownian motion.  However, there is still some question as to what the fractional exponent, $\alpha$, in the noise time correlation means.  The answer comes from the finance stochastic approach to fractional Brownian motion.

\section{Stochastic Process Approach}\label{sec:stoch-approach}
In stochastic language, regular Brownian motion, $B(t)$, is a continuous-time process with independent increments: a Wiener process \cite{bm-stochastics}.  A process of this kind has increments that are randomly distributed and scaled according to the size of the increments.  The number of increments in the process remains constant for depending on the time-scale of the process the size of the increments must be rescaled.  This random distribution is what leads the time correlation to be a Dirac delta.  However, in stochastics language, the time correlation is called the time covariance because it is the covariance of a motion at two points in time; from now on it will be referred to simply as the covariance.

\begin{figure}[t]
	\centering
			\includegraphics[width=4in]{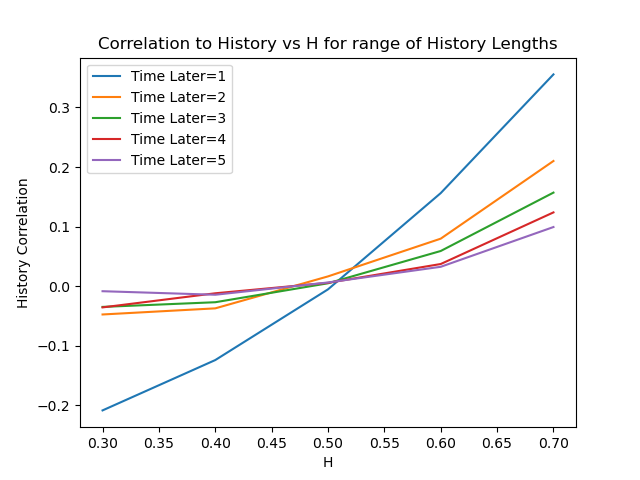}
		\caption{Using a time autocorrelation function, the curves here show the average strength of the historical correlation of a fractional Brownian motion path as a function of the Hurst parameter, $H$.  Each curve is the average correlation to time points a given number of steps into the future.}
		\label{fig:history-corr}
\end{figure}

In order to include memory in the system, these increments are changed such that successive increments are dependent on those that came before.  The parameter that defines the strength of that historical dependence is called the Hurst parameter, $H$, where $H$ = 1/2 is regular Brownian motion.  With this, the covariance of a fractional Brownian motion path, $B_H (t)$, is given by \cite{fBM-lang-connect}

\beq
\big\langle B_{H}(t) B_{H}(\tau) \big\rangle = \frac{\sigma_{0}^2 \Gamma(2-2H)}{4H\Gamma(3/2-H)\Gamma(1/2+H)} \bigg( t^{2H} + \tau^{2H} - \vert t-\tau \vert^{2H} \bigg),
\label{eq:covar-motion}
\eeq
where $\sigma_0$ is a constant that comes from regular Brownian motion.  From here on, this value is normalized out of the paths.  The Hurst parameter is limited to $H \in (0,1)$, and the question is what these different values of H actually correspond to.  Figure \ref{fig:history-corr} shows the time autocorrelation of fractional Brownian motion paths as a function of $H$.  The time autocorrelation measures the correlation between a point in the path at time $t$ with a point on the path some time later.  The correlation is averaged over points from the entire given path.  What is seen, in Figure \ref{fig:history-corr}, is that for $H$ = $1/2$ the correlation to history is zero, i.e. no memory.  For $H$ $<$ $1/2$, there is negative correlation to history while for $H$ $>$ $1/2$, there is a positive correlation.  From Ref \cite{fBM-lang-connect}, we know that this correlation as a function of the Hurst parameter is exactly correct.  Also, as expected, the strength of the correlation decays as the time separation gets larger.

However, it is important to know that motion is not the same as noise.  The increment process $\xi_H (t)=B_H (t+1) - B_H (t)$ is fractional Gaussian noise.  Now, because we are looking at a continuous-time process this expression needs to be continuous in time and so we get $\xi_H (t) = d B_H / dt$.  Above it was mentioned that $H$ = 1/2 is regular Brownian motion, so the noise generated by this motion is white noise.  The expression given in Equation \ref{eq:covar-motion} describes the covariance of the motion, not the noise.  The noise covariance is given by: 

\beq
\langle \xi_{H}(t) \xi_{H}(\tau) \rangle = \frac{ (2H-1)\Gamma(2-2H)}{2\Gamma(3/2-H)\Gamma(1/2+H)} \vert t - \tau \vert^{2H-2}.
\label{eq:noise-covar}
\eeq
The generalization of white noise generates what is known as coloured noise.  This power law relation also looks quite similar to Equation \ref{eq:noise-covar-lang}.

\section{Equivalence of Approaches} \label{sec:equiv-approaches}
The aim of the previous sections was to describe the two different approaches to generating fractional Brownian motion.  Theoretically, they both generate the same dynamics; hence they should be related to each other \cite{fBM-lang-connect}-\cite{mandelbrot-fbm}.  There is an obvious similarity in the power law form between Equations \ref{eq:noise-covar-lang} and \ref{eq:noise-covar}.  From the physics point of view, it is known that $f(t)$ is some kind of noise and from the financial stochastic point of view, it is known that $\xi_H (t)$ is noise.  Let's set them equal to each other, up to a constant $\epsilon$.  This means that we have the relation

\beq
\boxed{\alpha = 2-2H,}
\eeq
which is exactly the relation given in Ref \cite{fBM-lang-connect}.  Another point needs to be addressed as well: the phrase "up to a constant".  There are some coefficients in front of Equations \ref{eq:noise-covar-lang} and \ref{eq:noise-covar} that are non-trivial.  Taking the noise $f(t)$ to be the colored noise generated from a stochastic process means that the fractional Langevin equation from Equation \ref{eq:lang-3} becomes

\beq
M \frac{d^2x}{dt^2} + \gamma {}^{C}D_{0}^{\alpha} \big[ x(t) \big] = \epsilon \xi_H (t)
\label{eq:lang-4}
\eeq
along with the caveat of the coefficients:

\beq
\epsilon^2 = 2\gamma k_B T \frac{\Gamma(3/2-H)\Gamma(1/2+H)}{\Gamma(2H)\Gamma(2-2H)}.
\label{eq:fluc-dissip}
\eeq
In Section \ref{sec:lang-approach}, the word dissipation was used to describe the transfer of energy between the system and the bath and this was an intentional choice of words.  The coefficient $\epsilon$, maintains the fluctuation-dissipation theorem discussed in the Langevin case and connects to the stochastics approach as well.


In Ref \cite{robin-time-glass}, Laplace transforms were used to solve a fractional Langevin equation of the same form as Equation \ref{eq:lang-4}.  The analytical solution is:

\beq
x(t) = \frac{1}{M} \big[ \epsilon \xi_H(t) \big] \ast \bigg[ tE_{2H,2} \bigg( \frac{-\gamma}{M} t^{2H} \bigg) \bigg] + x'(0) tE_{2H,2} \bigg( \frac{-\gamma}{M} t^{2H} \bigg),
\label{eq:lang-soln}
\eeq
where $E_{a,b} (t)$ is the two-parameter Mittag-Leffler function, Equation \ref{eq:mitleff}, and it is taken that $x(0)=0$.

By combining what is known from the physics world of differential equations with the stochastic processes understanding of finance, we arrive at a fractional differential equation with coloured noise that describes fractional Brownian motion: a system with memory.

\chapter{Financial Model}
\label{chap:fin-model}
\section{Analytic Description}

Here, we give a bird's eye view of how a financial market can be modeled using the Langevin equation based on Ref. \cite{fin-lang} by Cont and Bouchaud.  It begins with the dynamics of simple supply and demand.  Assuming that supply and demand start off in equilibrium, for constant supply, when demand increases, the price increases because of the competition between actors.  Then, for constant demand, when supply increases, the price decreases because the value is tied to the availability.  Therefore, there exists a direct relationship between price and demand and an inverse relationship between price and supply.  For price, $x(t)$, demand $\phi_{D}$, and supply $\phi_{S}$,

\beq
\frac{dx}{dt} = u(t) = \frac{\Delta \phi}{\lambda}, \quad \Delta \phi = \phi_{D} - \phi_{S},
\eeq

\noindent where $u(t)$ are the returns and $\lambda$ is called the market depth.  The market depth is the amount of excess demand required to move the price by one unit, or in financial terms: market liquidity, from now on referred to just as liquidity.

\begin{table}
\centering
\begin{tabular}{ cc } 
Liquid & Illiquid \\  
\begin{tabular}{ |c|c|c| } 
\hline
 & \$1.20 (100)\\
 & \$1.19 (120)\\
 & \$1.18 (200)\\
 & \$1.17 (250)\\
 & \$1.16 (400)\\
 & \$1.15 (200)\\
\$1.14 (300) & \\
\$1.13 (450) & \\
\$1.12 (300) & \\
\$1.11 (280) & \\
\$1.10 (210) & \\
\hline
\end{tabular} & 
\begin{tabular}{ |c|c|c| } 
\hline
 & \$1.30 (2)\\
 & \$1.25 (5)\\
 & \$1.21 (10)\\
 & \$1.18 (7)\\
 & \$1.16 (2)\\
 & \$1.14 (1)\\
\$1.10 (5) & \\
\$1.07 (10) & \\
\$1.02 (15) & \\
\$0.95 (4) & \\
\$0.91 (2) & \\
\hline
\end{tabular} \\
\end{tabular}
\caption{For each price, which is denominated in dollars, there is an associated number of orders that could be satisfied by either buying or selling at that given price.  The number of orders is in the parentheses next to the price.  The majority of financial markets today are traded using orderbooks of this form.}
\label{tab:liquidity}
\end{table}

In a financial market, there is an orderbook: a list of all prices that the actors are willing to buy and sell the asset.  When that orderbook only has a small number of actors willing to buy/sell the asset at only a few prices, as in the right side of Table \ref{tab:liquidity}, the market is illiquid.  When large transactions are made, the orderbook will empty out and the current price of the asset will change drastically.  When the orderbook is well filled, as in the left side of Table \ref{tab:liquidity}, there are large numbers of actors willing to buy and sell at any price around the current value: high liquidity.  This means that even when a transaction is made that buys/sells a large amount of the asset, the orderbook stays well filled and mostly unchanged.

Intuitively, the market depth, $\lambda$, is analogous to mass.  Large masses have a lot of inertia and thus are less subject to changes in their position.  This is mirrored in $\lambda$, where substantial market depth means that the price is less subject to change.

Market makers are the entities that satisfy orders for actors, and the rate at which they can fulfill those orders is given by the parameter $\Gamma$, which is large for highly liquid markets.  We assume that market makers act symmetrically such that they complete buy orders at the same rate as sell orders.  This allows us to expand $\Gamma$ as

\beq
\Gamma (\phi) = \beta + \beta' \phi + ...
\eeq
where $\beta$ is a lowest-order approximation to the rate at which market makers satisfy orders.  The equation for the market makers is then:

\beq
\frac{d\phi_{D,S}}{dt} \bigg\vert_{MM} = -\Gamma(\phi_{S,D}) \phi_{D,S},
\eeq
\beq
\frac{d \Delta \phi}{dt} \bigg\vert_{MM} = \lambda \frac{d^2 x}{dt^2} = -\beta \big( \phi_D - \phi_S \big) = - \beta \lambda \frac{dx}{dt}.
\eeq
Now we have a differential equation that describes the change in price given that the market makers process transactions at the rate $\Gamma$.

The change in price of an asset, from the actor point of view, can be modeled as a function of two terms: a random term, $f_{D,S}$(t), and a trend term, $m_{D,S}$(t).

\beq
\frac{d \phi_{D,S}}{dt} \bigg\vert_A = m_{D,S} (t) + f_{D,S} (t).
\eeq
The trend term is a function of the anticipated return, R(t), and the anticipated risk, $\Sigma$(t), for demand and supply.  In this particular case, we assume that the asset is risk neutral, which is a first-order approximation in the returns that assumes the risk to be constant in time: $\Sigma$(t)=$\Sigma^0$.  The anticipated return is given by

\beq
R_{D,S}(t) = R_{D,S}^{0} + a_{D,S} \int_{0}^{t} d \tau K_R (t-\tau) \frac{dx}{d \tau},
\eeq
where $a_{D,S}$ measures the impact of the observed recent trend on the anticipated return and $K_R$ is a normalized kernel that describes how the historical trend is measured by the users.  We assume that the asset price has no general trend up or down, so the price is only determined by the demand/supply interaction and the random term.  This mean-reverting behavior is a reasonable assumption because the system here is isolated and thus is not subject to external factors.    It then follows that the constant terms in the risk and return cancel: $R_{D,S}^{0}$=$\Sigma_{D,S}^{0}$.

The form of the anticipated return looks very similar to a fractional derivative.  Indeed if we take

\beq
K_R(t)=\frac{t^{-\alpha}}{\Gamma(1-\alpha)},
\eeq
we get the Caputo fractional derivative of price.  Putting together the return and risk for demand and supply, we get:

\beq
\frac{d \Delta \phi}{dt} \bigg\vert_{A} = \frac{(a_D - a_S)}{\Gamma(1-\alpha)} \int_{0}^{t} d \tau (t-\tau)^{-\alpha} \frac{dx}{dt} + \big( R_{D}^{0} - \Sigma_{D}^{0} \big) - \big( R_{S}^{0} - \Sigma_{S}^{0} \big) + f(t), \quad \textrm{and}
\eeq
\beq
\lambda \frac{d^2 x}{dt^2} \bigg\vert_A = a D^{\alpha} \big[ x(t) \big] + f(t),
\eeq
where we use the null-trend condition discussed above to move to the second line, $f(t)$ = $f_D$(t) - $f_S$(t), and $a$ = $a_D - a_S$.

To get the full stochastic differential equation for price, we combine the market maker component and the actor component to get:

\beq
\lambda \frac{d^2 x}{dt^2} + \beta \lambda \frac{dx}{dt} - a D^{\alpha} \big[ x(t) \big]  = f(t).
\label{eq:fin-lang-first}
\eeq
In the limiting case of no memory, Ref \cite{fin-lang} shows that the fractional derivative term in Equation \ref{eq:fin-lang-first} just goes to a first order derivative like a regular friction term.  Therefore, we take $\alpha$ = 1 as an upper bound and allow $\alpha$ to range in $(0,1)$.  For $\alpha$ in this range, that means that the Hurst parameter, $H \in (1/2,1)$.  As discussed in Section \ref{sec:stoch-approach}, when $H$ $>$ $1/2$, the path has a positive correlation with history.  From a financial point of view this actually makes a lot of sense.  If $H$ $<$ $1/2$, then there would be a negative correlation with history meaning that when the price goes up, actors expect the price to go down next.  This idea is quite contrary to the fundamental behavior of financial systems so it is reasonable to restrict $\alpha$ to this range.

\subsection{Noise and the Fluctuation-Dissipation Theorem}
\label{sec:thermal-energy}

Ref \cite{fin-lang} also gives the covariance of the random term as a function of the market depth, $\lambda$, and a parameter $\Theta$:

\beq
\langle f(t) f(t') \rangle = 2 \lambda^2 \Theta \delta (t-t')
\eeq
where $\Theta$ is the susceptibility of the market to random fluctuations.  The whole aim of our description here is to look at the impact of history, and in principle, $\Theta$ should be dependent on history.  If the system was hit by a large event recently, actors are on edge and will be more reactionary to incoming external news.  So, the covariance will be,

\beq
\langle f(t) f(t') \rangle = 2 \lambda^2 \big[ \Theta_1 \delta (t-t') + \Theta_2 (t-t')^{-\alpha} \big].
\eeq
Therefore, the random term, $f(t)$, can be thought of as the sum of two, uncorrelated, noise terms.  One of them is white noise, which leads to the $\delta(t-t')$, and the other is a colored noise term which leads to the $(t-t')^{-\alpha}$ because of the memory.

Mathematically, this also makes sense when we look back at Equation \ref{eq:fin-lang-first}, which has an integer-order derivative and a fractional-order derivative.  Using the fluctuation-dissipation theorem, the integer-order term interacts with the white noise and the fractional-order term interacts with the colored noise.  Then, we notice that the fractional-order term is negative. Thus, the random term $f(t)$ should be written as

\beq
f(t) = \eta(t) - \xi_{\alpha} (t),
\eeq
where $\eta$ is the white noise and $\xi_{\alpha}$ is the colored noise.  Both of the noises include constants that satisfy the fluctuation-dissipation theorem: $b_W$ and $b_C$.  Using all of these relationships, we find the following relations between the constants,

\beq
\Theta_1 = \frac{\beta k_B T}{\lambda}, \quad b_{W}^{2} = 2 \beta \lambda k_B T; \qquad \Theta_2 = \frac{a k_B T}{2 \lambda^2 \Gamma(1-\alpha)}, \quad b_{C}^{2} = a k_B T \frac{2 \Gamma((1+\alpha)/2) \Gamma((3-\alpha)/2)}{\Gamma(2-\alpha)\Gamma(\alpha)}.
\eeq
These coefficients all include a factor of $k_B T$ which in physics language is the thermal energy of the system.  In physics, thermal energy corresponds to the average velocity of particles within a system.  In a financial system, position can be thought of as the amount of an asset an actor has and thus velocity is how much the actor buys/sells the asset.  The thermal energy then could be related to the trading volume at a given time.  High temperature financial systems would be ones that are traded frequently while low temperature ones correspond to quieter systems.

With these two components, we essentially have a particle, our price, interacting with two different, uncorrelated, baths.  The first bath has no memory and is just pushing the price around based on global market expectations.  The second bath is the fractional Brownian motion which is essentially the individual users interacting with the price in a way that is inspired by their memory.  Hence, our full fractional stochastic differential equation is

\beq
\lambda \frac{d^2 x}{dt^2} + \beta \lambda \frac{dx}{dt} - a D^{\alpha} \big[ x(t) \big] = \eta(t) - \xi_{\alpha} (t),
\label{eq:fin-lang-eq}
\eeq
An interesting side-note would be that the values for $\Theta_1$ and $\Theta_2$ are given by the fluctuation-dissipation theorem and are based on the financial system responding to noise in an expected manner.  It would be interesting to investigate an out-of-equilibrium system when those values for $\Theta_1$ and $\Theta_2$ do not hold, especially given the work in Ref. \cite{robin-time-glass} on out-of-equilibrium systems.

\subsection{Financial Intuition from the Limiting case of No Memory}
Now, instead of taking the kernel $K_R$ to obtain the Caputo fractional derivative, we consider that the kernel has no memory, then the fractional derivative goes to integer order and the differential equation becomes

\beq
\lambda \frac{d^2 x}{dt^2} + (\beta \lambda - a) \frac{dx}{dt} = f(t),
\label{eq:no-memory}
\eeq
where according to the fluctuation-dissipation theorem and stable solution existence, $\beta \lambda \geq a$.  The left-hand side is the rate of change of demand when the price moves a single unit which is a dynamical variable describing the entire market.  This describes how the demand is expected to change in time when the price moves a single unit based purely on the market depth.  When the relationship is strictly equal, it would mean that the impact of recent trends on user's actions exactly matches up with how the market expects users to respond.  This is an entirely capitalistic model in which users exclusively seek profits.  The friction term goes to zero and we are left with a simple stochastic differential equation, but this is not interesting for us.

When $\beta \lambda < a$, the impact of recent trends is less than what the market expects.  An idealistic trader in the market will hold onto the asset independent of how the price is behaving because they are a core believer in whatever the asset represents.  A timely example of this would be Bitcoin: there exists Bitcoin users who believe so strongly that it is the future so no matter what the price is, they will never sell.  The impact of recent trends is essentially averaged over all users in the market so if there are idealists in the market, they do no contribute to the parameter $a$ and thus the value is less than what the market depth would predict.  Therefore, the existence of idealists in the market is what generates friction in this model.  The term used here is idealists, but that implies an assumption of their motivation.  These actors can also been seen as non-optimal users in the market, i.e., a kind of disorder.

\chapter{Numerical Simulation of Motion}
\label{chap:num-sim}
Now that we have an analytical description of our model, it needs to be tested to see how exactly it behaves.  The nature of fractional Brownian motion is stochastic, so we numerically calculate paths and typically do an ensemble average to see the dynamics.
\section{Colored Noise}
In order to simulate fractional Brownian motion, we use a method from Ref \cite{fBM-sim-meth2} which goes as follows.  Given a regular Brownian motion, $B(t)$, and Hurst parameter, $H$, a fractional Brownian motion path can be written as

\beq
B_{H} (t) = \int_{0}^{t} K_{H} (t,s) dB(s), \qquad K_H (t,s) = \frac{(t-s)^{H-1/2}}{\Gamma (H+1/2)} \prescript{}{2}{F}_1 \bigg( H-\frac{1}{2}; \frac{1}{2}-H; H+\frac{1}{2}; 1-\frac{t}{s} \bigg),
\eeq

\noindent where $\prescript{}{2}{F_1}$ is the Euler hypergeometric integral.  In order to do this numerically, the integral is turned into a sum given by:

\beq
B_{H} (t_j) = \frac{n}{T} \sum_{i=0}^{j-1} \Delta B_{i+1} \int_{t_i}^{t_{i+1}} K_H (t_j,s) ds,
\eeq

\noindent where $n$ is the number of steps between $t=0$ and $t=T$, and $\Delta B_{i}$ is a regular Brownian motion increment selected from $\sqrt{\Delta t}\cdot\mathcal{N}(0,1)$.  The integral in this equation is calculated numerically using Gaussian quadrature.
\subsection{Qualitative Analysis}
\begin{figure}[t]
	\centering
			\includegraphics[width=4in]{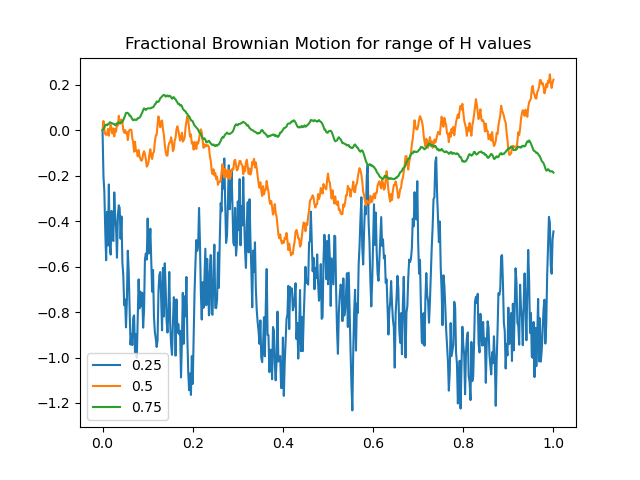}
		\caption{This plot shows three different fractional Brownian motion paths made using a range of $H$ values.  The simulation was done for 500 steps within a time interval in (0,1) seconds.}
		\label{fig:fBM-meth2}
\end{figure}

Figure \ref{fig:fBM-meth2} shows this simulated motion for three values of the Hurst parameter.  From the literature \cite{fBM-lang-connect} and \cite{mandelbrot-fbm}, it is known that a fractional Brownian motion path with H $>$ 1/2 is positively correlated with history and thus is smoother.  When the motion has H $<$ 1/2, the path is negatively correlated with history, hence the paths are more volatile.  These qualitative characteristics are generated by our simulation as shown in Figure \ref{fig:fBM-meth2}.  When a given path has a larger standard deviation than another, the first path is said to be more volatile relative to the second.

\subsection{Quantitative Analysis}
For a fractional Brownian motion path, two values are calculated to assess the accuracy of a simulated motion \cite{fBM-lang-connect}.  The first is:

\beq
E \big[ B_{H}^{2} (t=1) \big] = \frac{ \Gamma(2-2H)}{2H \Gamma(3/2-H) \Gamma(1/2+H)}.
\label{eq:b2}
\eeq

\begin{figure}
     \centering
     \begin{subfigure}[b]{0.45\textwidth}
         \centering
         \includegraphics[width=\textwidth]{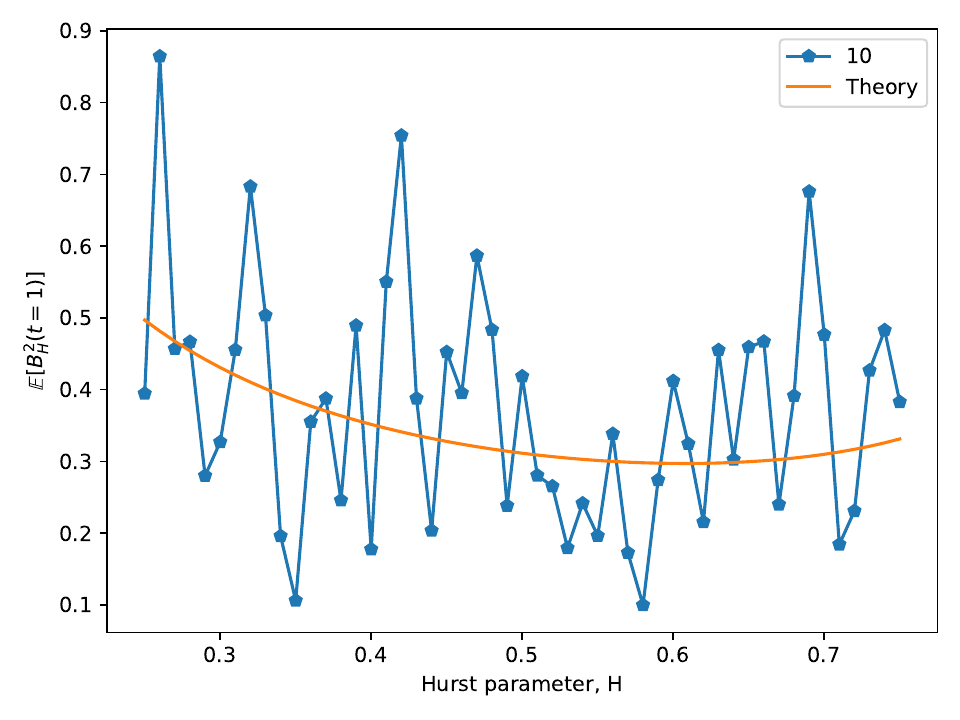}
         \subcaption{}
     \end{subfigure}
     \hfill
     \begin{subfigure}[b]{0.45\textwidth}
         \centering
         \includegraphics[width=\textwidth]{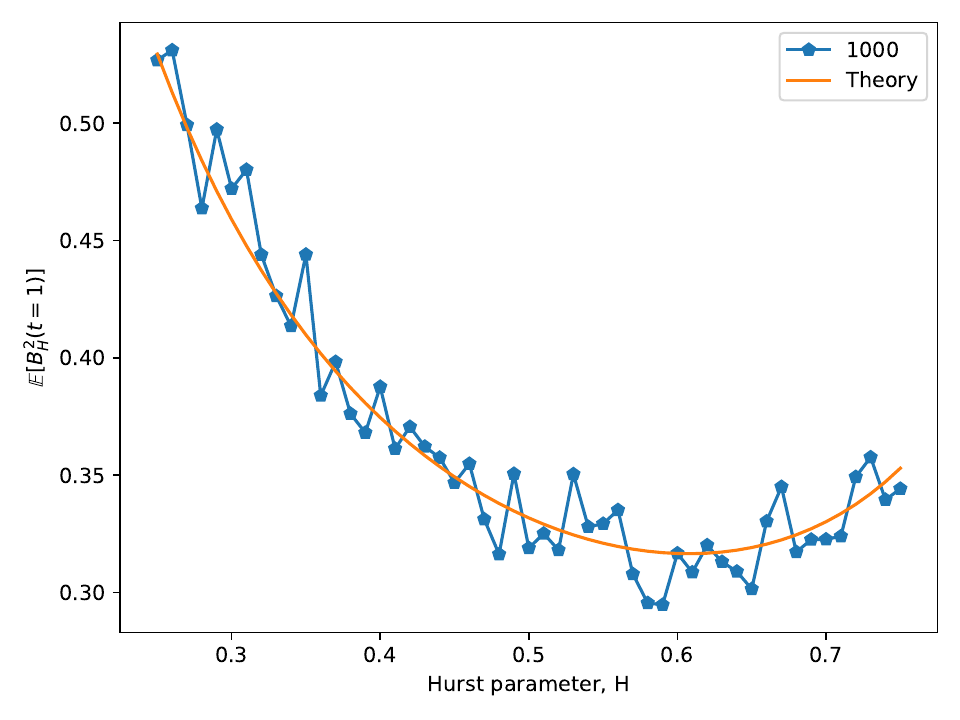}
         \subcaption{}
     \end{subfigure}
        \caption{The figures here show the numerical calculation, the blue dotted lines, of Equation \ref{eq:b2} for two different numbers of average runs: 10 runs in (a) and 1000 runs in (b).  These are compared with the theoretical predictions: the orange curves.}
        \label{fig:b2-hurst-2}
\end{figure}

Figure \ref{fig:b2-hurst-2} shows the calculated behavior along with the theoretical behavior of the expected value for a range of Hurst parameters for two different numbers of averaging runs.  The right figure with 1000 averaging runs has much less noise.  As the number of averaging runs increases, the curve converges towards the theoretical expectation which provides confidence in the accuracy of the fractional Brownian motion generation method.  It is also interesting to investigate the behavior of the time-separated covariance of the motion:

\beq
\textrm{Cov}\big( B_{H}(t=\textrm{end}) B_{H}(s) \big) = \frac{ \Gamma(2-2H)}{4H \Gamma(\frac{3}{2}-H) \Gamma(\frac{1}{2}+H)} \bigg( t^{2H} + s^{2H} - \vert t-s \vert^{2H} \bigg).
\eeq

\begin{figure}
     \centering
     \begin{subfigure}[b]{0.45\textwidth}
         \centering
         \includegraphics[width=\textwidth]{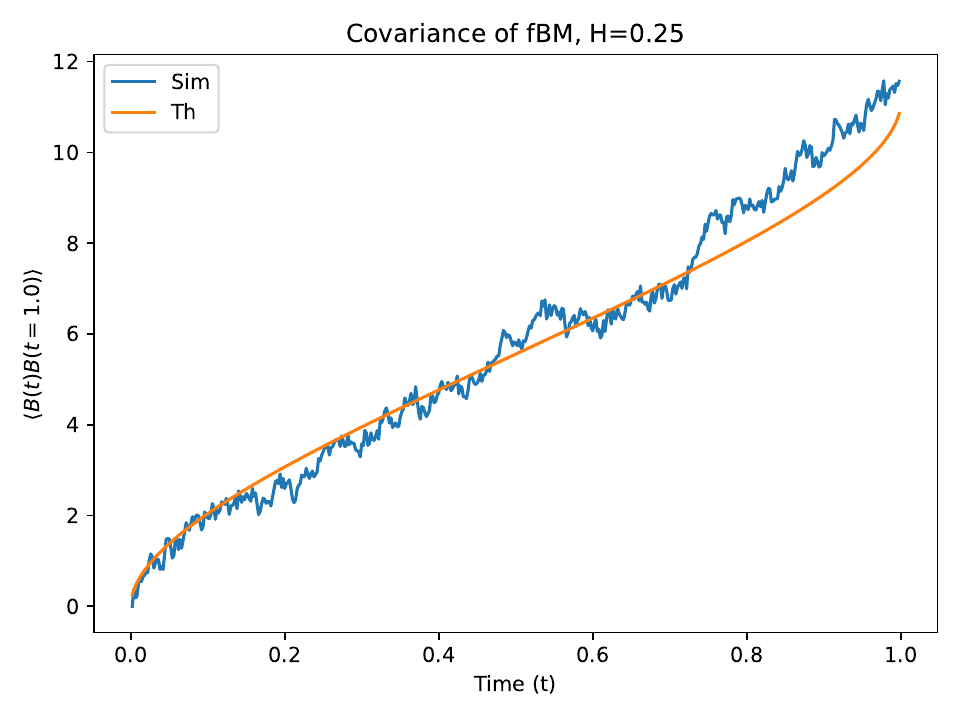}
         \subcaption{}
     \end{subfigure}
     \hfill
     \begin{subfigure}[b]{0.45\textwidth}
         \centering
         \includegraphics[width=\textwidth]{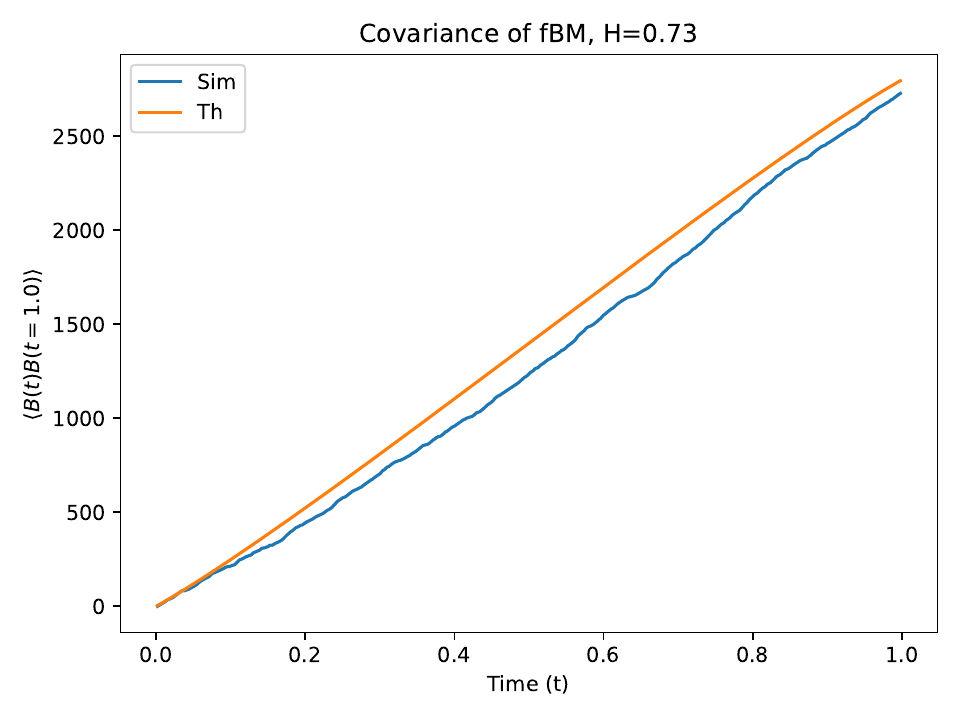}
         \subcaption{}
     \end{subfigure}
        \caption{These curves show the covariance for two values of the Hurst parameter: (a) $H=0.25$, (b) $H=0.73$.  The covariance here is calculated with respect to the end point of the simulation with 500 time steps and compared to the theoretical expectation in orange.}
        \label{fig:covar}
\end{figure}

\noindent Figure \ref{fig:covar} shows the theoretical and calculated covariance for different values of the Hurst parameter.  The inaccuracy in the two values as compared to the theory comes from two factors.  The first is that the expected value is calculated averaging over some number of runs and the values become more accurate as the number of runs increases.  In order to generate Figure \ref{fig:covar}, the values were averaged over 200 runs.  The curves were also generated for smaller numbers of averaging runs and convergence towards the theoretical value was clearly seen in the same way as Figure \ref{fig:b2-hurst-2}.  The other source of inaccuracy comes from the size of the time steps that are taken: Figure \ref{fig:covar} was calculated with $0.002$ as the time step.  As the time step gets smaller, the accuracy of the covariance curve increases.

\section{Finding Numerical Solutions to Differential Equations}

At this point, we have a differential equation that describes a financial system with memory and a numerical method for generating fractional Brownian motion.  Now, we solve the equation.  Equation \ref{eq:fin-lang-eq} is quite difficult to solve analytically, so instead we decide to solve it numerically.  The key to most numerical equation solvers is how the next guess at the solution is made.  The method of choosing the next guess is typically built around what kind of equation is trying to be solved.  However, this can require extensive knowledge of the equation being solved.  For some ease in building a numerical solver, we use a simple Metropolis-Hastings Monte Carlo method to find our next guess \cite{mc-metro}.  The method goes as follows,

\begin{multicols}{2}
\textbf{Monte Carlo Pseudo-code}
\begin{small}
\begin{enumerate}
    \item Start with function, 
    
    $g(t) = \lambda \frac{d^2 x}{dt^2} + \beta \lambda \frac{dx}{dt} - \eta(t) + y_H (t)$

    \item Find first guess, $g(t)=0$
    \item Calculate residuals,
    
    $res^{(i)} = \bigg\vert g(t; x^{(i)}) (t) - a D^{\alpha} \big[ x^{(i)} (t) \big] \bigg\vert$

    \item If $res^{(i)} >$ tolerance for all $i$, guess new solution. Else, solution is found.
    \item Continue for $n_{\textrm{max}}$ steps.
\end{enumerate}
\end{small}
\columnbreak
\textbf{Next Guess Pseudo-code}
\begin{small}
\begin{enumerate}
    \item Make list of all time points with $res^{(i)} >$ tolerance. Select first four.
    \item For each of the four selected time points
    \begin{enumerate}
        \item Generate random movement: either up or down, with maximum step size
        \item If new price decreases residuals, accept move.  Else, reject move.
    \end{enumerate}
    \item Check if $res^{(i)} <$ tolerance for all time points
\end{enumerate}
\end{small}
\end{multicols}
The function $g(t)$ is the full differential equation without the fractional derivative term.  Thus, it is straight-forward to find an analytic solution.  The interesting part is to figure out how to guess the next possible solution.  This is where the Monte Carlo method comes in.

The residue for each time point is taken to effectively be the energy of that point.  A Monte Carlo method works by minimizing the energy.  The algorithm proposes a random change in the price position at that time point, with a given hard-coded maximum change, and calculates the new residues for all points in time.  If the new price position generates a residue that is smaller than those of the previous time point, then the new price position is accepted.

Normally, a Monte Carlo code would attempt to move every single time point at each Monte Carlo step, however our code does not do this.  The residuals calculate how close a given time point is to the true solution and because they depend on history, the residuals of a given time point depend on the time points in the past.  If the time points in the past are not within the prescribed tolerance to the true solution, then the residuals calculated do not accurately represent the closeness of this time point to its true solution.  An attempt to move this time point would then be useless because its acceptance does not guarantee that it is now closer to the solution.

To avoid this issue, moves are only attempted for a small group of time points, as seen in the \textbf{Next Guess Pseudo-code}.  In this way, the solution is found by effectively moving through time so that the history used to calculate the residuals is actually representative of the true history of a given time point.



In order to prove that this numerical method in fact does find the proper solutions, it is tested on two different differential equations.  Proving that the method works to find true solutions for two systems provides strong confidence that it will also find the solution to a system for which we do not know the solution.

\subsection{Simple Coloured Noise Force}
The first system that we look at is a rather simple one:

\beq
\frac{d^2 x}{dt^2} = \xi_H (t),
\label{eq:simp-noise-diffeq}
\eeq
where $\xi_H$ is coloured noise.  This is in essence a particle being acted upon by a coloured noise force.  The $g(t)$ for this equation is simply $g(t)=d^2 x/dt^2 - \xi_H (t)$ and the residuals are just the magnitude of $g(t)$.


\begin{figure}[t]
	\centering
			\includegraphics[width=4in]{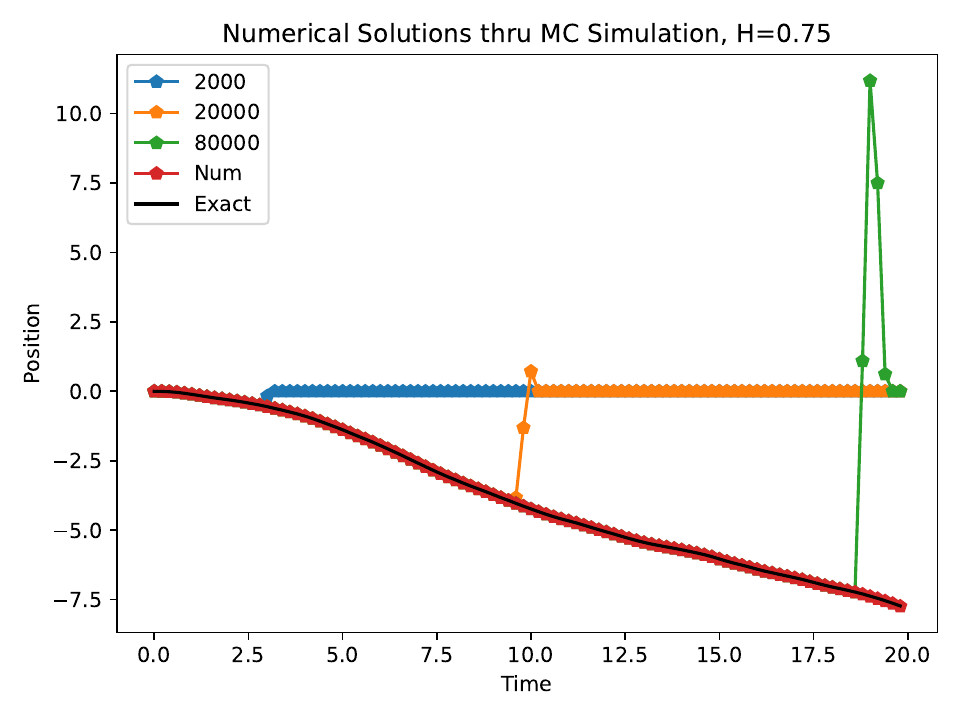}
		\caption{The black curve shows the exact solution to Equation \ref{eq:simp-noise-diffeq} over 20 seconds.  The other lines show the estimated solution at a given number of Monte Carlo steps.  As the number of steps increases, more and more of the curve overlaps with the theoretical prediction.  The initial guess was all zeroes.}
		\label{fig:just-noise-solns}
\end{figure}

The algorithm is essentially trying to minimize the residues for all points in time until they are smaller than the given tolerance.  Figure \ref{fig:just-noise-solns} shows the solution over time for a few Monte Carlo time slices.  The plot shows how the solution is found by moving through time.  The first guess is all zeros, and then small groups of time points are brought towards the solution by the Monte Carlo simulation.


\subsection{Fractional Langevin Equation}

We have shown that our Monte Carlo method works for a simple second-order differential equation.  Now we verify whether it works for a fractional differential equation, specifically Equation \ref{eq:lang-4}.  With this system, $g(t)$ = $M \frac{d^2 x}{dt^2} - \eta \xi_H (t)$ and the residuals function is the magnitude of $g(t) + \gamma {}^{C}D^{\alpha} x(t)$.

In the definitions of different fractional derivatives, we said that the Riemann-Liouville and Caputo forms both are numerically divergent and therefore useless in this Monte Carlo method.  That means that for the simulation we need to use the Gr\"uwald-Letnikov method from Equation \ref{eq:gru-let}.  In this case, we are taking $x(0)$ = 0, which means that, from Equation \ref{eq:trans-rl-2-cap}, all three derivative forms are equivalent.

\begin{figure}
     \centering
     \begin{subfigure}[b]{0.45\textwidth}
         \centering
         \includegraphics[width=\textwidth]{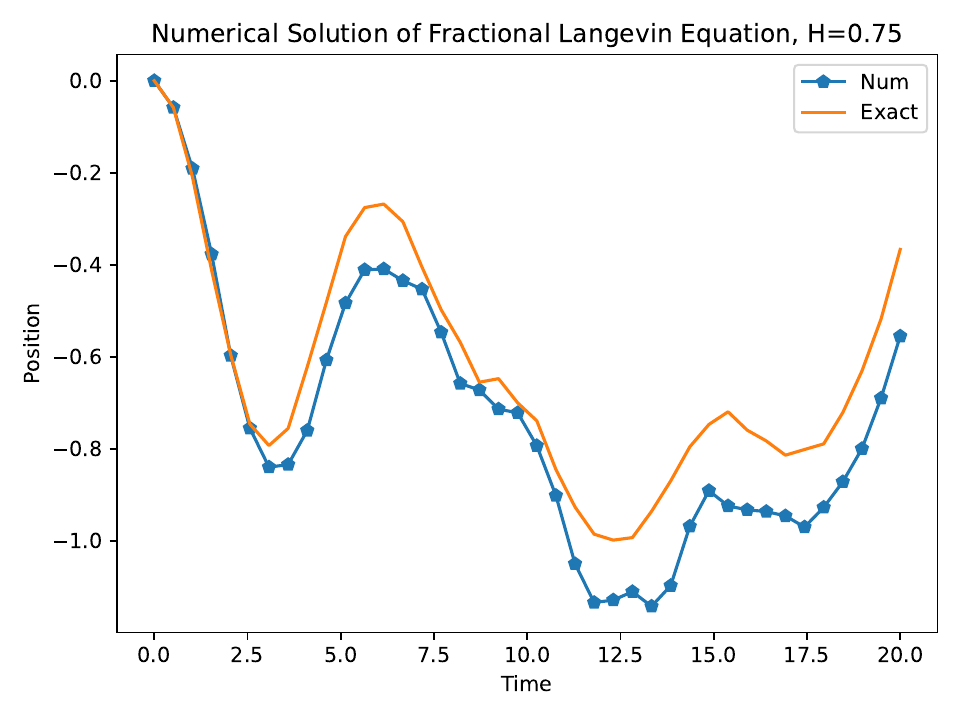}
         \subcaption{}
     \end{subfigure}
     \hfill
     \begin{subfigure}[b]{0.45\textwidth}
         \centering
         \includegraphics[width=\textwidth]{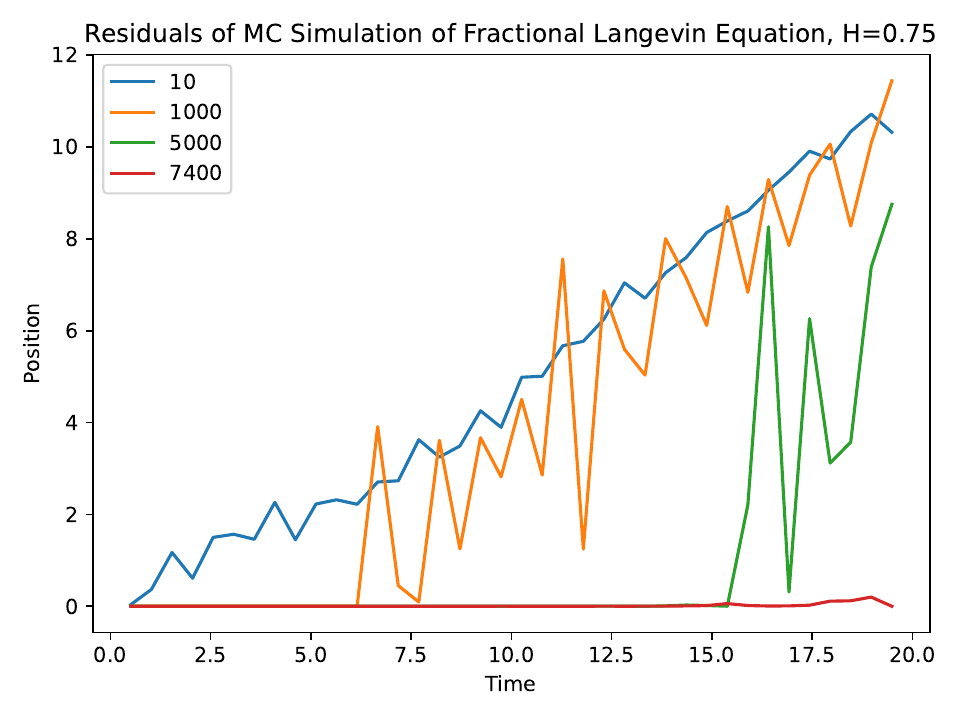}
         \subcaption{}
     \end{subfigure}
        \caption{Figure (a) shows the numerically calculated solution to Equation \ref{eq:lang-4} as compared to the analytical solution from Equation \ref{eq:lang-soln} for $H$ = 0.75 over 20 units of time.  The constants are taken to be $M=1=\gamma=k_B T$.  Figure (b) shows the residuals of the solution guess over the Monte Carlo simulation.}
        \label{fig:fraclang-num}
\end{figure}

Looking at the Figure \ref{fig:fraclang-num}(a), we observe that the numerical solution is not exactly what is expected given the known analytic solution.  From Figure \ref{fig:just-noise-solns}, the Monte Carlo method does find exact solutions and from Figure \ref{fig:frac-derivs-num}, the numerical calculation of the fractional derivative is precise and accurate for short times.  It turns out that the exact solution to the fractional Langevin equation is not a minimum of the residuals in the code we have used.  The assumption is that this has to do with the discrete nature of the simulation.  As the number of time points increases, more and more points align with the exact solution as the numerical solution becomes more accurate.

However, the shape of the numerical solution does still rather closely follow that of the exact solution.  Thus, it is reasonable to use this method of finding numerical solutions to examine the shape of the solutions for differential equations without analytical solutions.

\subsection{Optimization}
The key attribute of this project is the inclusion of memory in a stochastic system.  This impacts the dynamics but it also impacts the numerical method of finding solutions.  Using a Monte Carlo method to solve an equation with memory is a little bit like playing Whack-A-Mole.  When one time point is moved in a beneficial manner, the residuals of a time point in the future may increase: one mole goes down, another one pops up.  The inclusion of memory impacts the convergence properties of a given algorithm.  So, to understand this Monte Carlo algorithm, we look at the fractional Langevin equation.

\begin{figure}[t]
	\centering
			\includegraphics[width=4in]{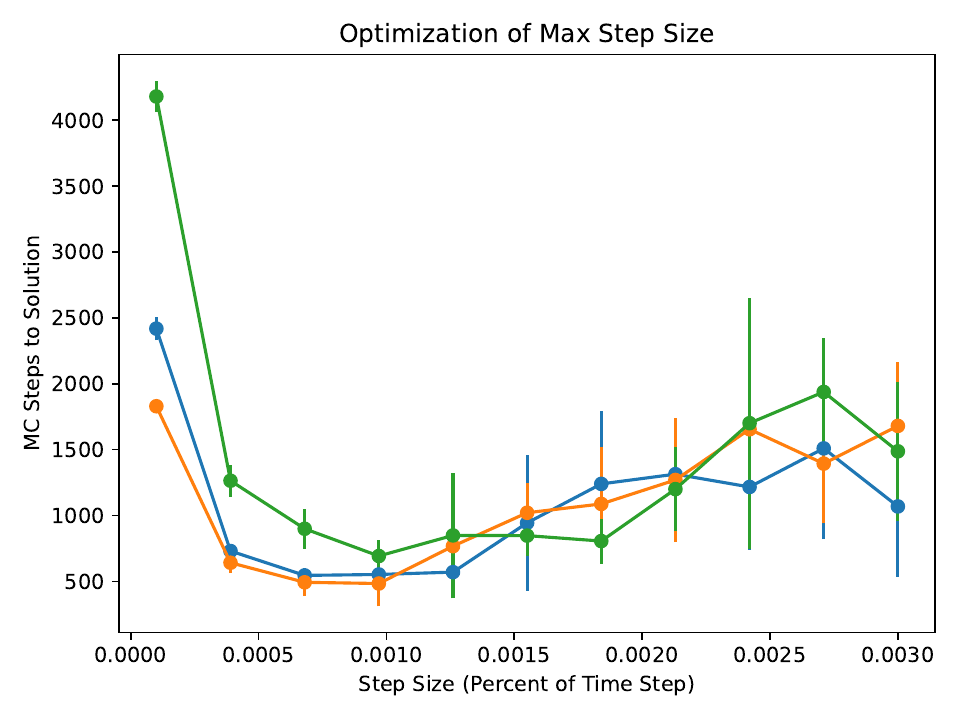}
		\caption{The optimization procedure is run three separate times and each curve comes from those three separate iterations.  Each point is gathered from an average of 5 attempts to find a solution and the error bars are the standard deviation of that set of Monte Carlo steps.}
		\label{fig:optim-stepsize}
\end{figure}

Figure \ref{fig:optim-stepsize} shows the average number of Monte Carlo steps to find a solution of the fractional Langevin equation with an accuracy smaller than $1\%$.  This particular plot was made for H = 0.75 and 10 time points in $t \in (0,1)$.  The different curves come from three different noises that all have the same Hurst parameter.  The plot shows that a maximum step size of about $0.1 \%$ of the time step minimizes the number of Monte Carlo steps it takes to find a solution.  

\chapter{Simulation of the Financial System}
\label{chap:analysis}
\begin{figure}[b!]
	\centering
			\includegraphics[width=4in]{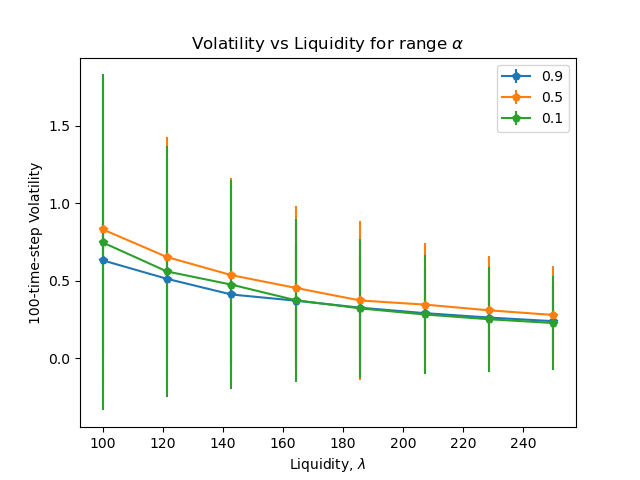}
		\caption{The plot shows the finite time realized volatility as a function of the market liquidity $\lambda$ for three values of the fractional derivative order.  The large error bars show that the liquidity-volatility relationship is independent of the memory in the system.}
		\label{fig:volat-liquid}
\end{figure}
In order to analyze the behavior of our financial model we examine the Mean Squared Displacement (MSD),

\beq
\textrm{MSD}(t) = \big\langle \vert x(t) - x(t=0) \vert^2 \big\rangle,
\eeq
where $\langle...\rangle$ is an ensemble average over a given number of different paths.  The MSD is effectively equivalent to the variance and thus can be used to calculate volatility and to examine the dispersion relations of our system.

Unless otherwise specified, the system begins with an initial velocity which is related to the thermal energy in the system.  This is done in order to start-off the system in thermal equilibrium because we are assuming that the system always follows the equipartition theorem.  Therefore, the initial velocity is always $v_0=\sqrt{k_B T/\lambda}$.

\section{Liquidity and Volatility}
As previously discussed, the liquidity of a given financial market plays an important role in the volatility of that market.  Therefore it is important that our model reproduces these same relationship.  One definition of volatility is the variance over a finite period of time, and is because the MSD is equivalent to the variance, the MSD can be used to calculate volatility.

Figure \ref{fig:volat-liquid} shows the 100-time-step volatility as a function the market liquidity parameter, $\lambda$, for a few values of the fractional derivative order $\alpha$ where all other parameters are kept the same.  As expected, as the market liquidity increases, the finite time volatility decreases showing that more liquid markets are more stable than less liquid markets.  The error bars here are quite large because the volatility comes from the MSD which is only averaged over 20 paths.  This comes from the computational limits of our simulation.  However, the trend between liquidity and volatility is still clear enough to say the relationship does exist.


However, the large error bars suggest that is no relationship between the volatility and the fractional derivative order.  Indeed, the existence of memory in the system does not play a significant enough role to change the fundamental relationship between volatility and liquidity.

\section{Expected MSD Behavior}
For all MSD plots, the parameters being used are $\lambda=500$, $\beta=1.0=a$ and the time scale is in seconds. To understand the asymptotic behavior of the equation \ref{eq:fin-lang-eq}, we look at four different interaction components: fractional derivative with colored noise, fractional derivative with white noise, integer derivative with colored noise, and integer derivative with white noise.  From Section \ref{sec:thermal-energy}, we know that the fractional derivative with colored noise and the integer derivative with white noise are governed by the fluctuation-dissipation theorem.

In the limit of no memory, the fractional derivative with colored noise can be neglected, and the regular integer order Langevin equation is retrieved, which leads to regular Brownian motion.  The noise term on the right-hand side is white noise and the friction term is a first order derivative.

From Lutz \cite{lutz}, when the Langevin equation contains just a fractional derivative and colored noise term, the scaling goes as $t^{\alpha}$ in the long-term.  The limiting case for this where $\alpha \rightarrow 1$ also maintains the linear scaling of regular Brownian motion.

However, because our system contains both a fractional derivative and an integer derivative, there exists more complex interactions between these terms. From the master's thesis of Robin Verstraten, \cite{robin-thesis}, we also know what happens with a fractional derivative interacting with a white noise term.  The interaction is out-of-equilibrium behavior which does not follow the fluctuation-dissipation theorem, but instead has meta-stable states as a function of the fractional derivative order, $\alpha$.  The long-term scaling of the MSD then goes as $t^{2\alpha-1}$ which is why it is shown for comparison in Table \ref{fig:all-msds}.

\section{Dispersion Regions}
\begin{figure}[h]
     \centering
     \begin{subfigure}[b]{0.48\textwidth}
         \centering
         \includegraphics[width=\textwidth]{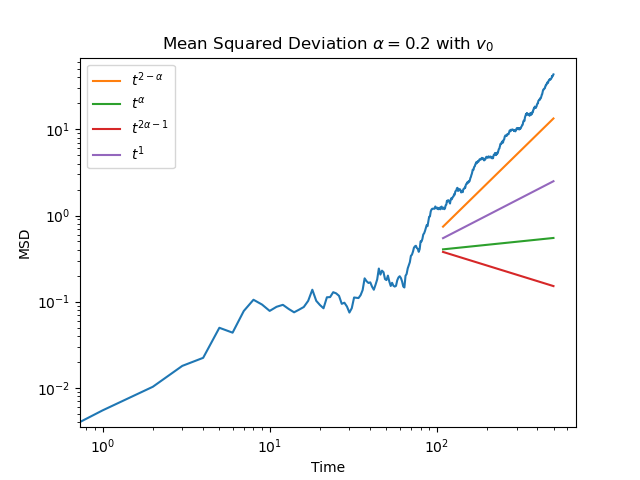}
         \subcaption{}
     \end{subfigure}
     \hfill
     \begin{subfigure}[b]{0.48\textwidth}
         \centering
         \includegraphics[width=\textwidth]{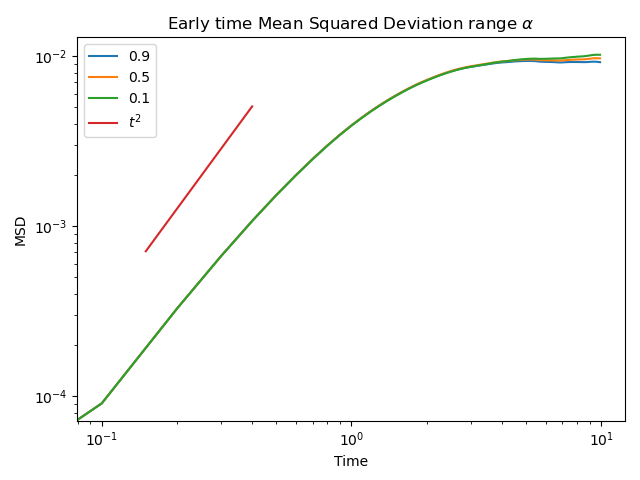}
         \subcaption{}
     \end{subfigure}
        \caption{The MSD here is calculated by an ensemble average over 20 different paths which each have different uncorrelated white and colored noise terms.  Each path is a solution to Equation \ref{eq:fin-lang-eq}.  The other lines are references to examine the long-term scaling of the dispersion.  Plot (a) shows the MSD for $\alpha$ = 0.2 up to 500 seconds.  Plot (b) shows the MSD for the first 10 seconds for $\alpha$ = 0.9, 0.5, and 0.1.}
        \label{fig:msd-0p2}
\end{figure}

Figure \ref{fig:msd-0p2}(a) shows the MSD for a fractional derivative order of $\alpha=0.2$, which has a positive correlation to history, and a finite initial velocity.  The figure shows three distinct regions each with different dispersion relations.  At the start, the curve goes as $t^2$ which is ballistic dispersion.  This can be seen more readily in Figure \ref{fig:msd-0p2}(b).  These are the times where the trace "particle" is exploring the environment and does not yet have any memory.

\begin{figure}[h]
	\centering
			\includegraphics[width=4in]{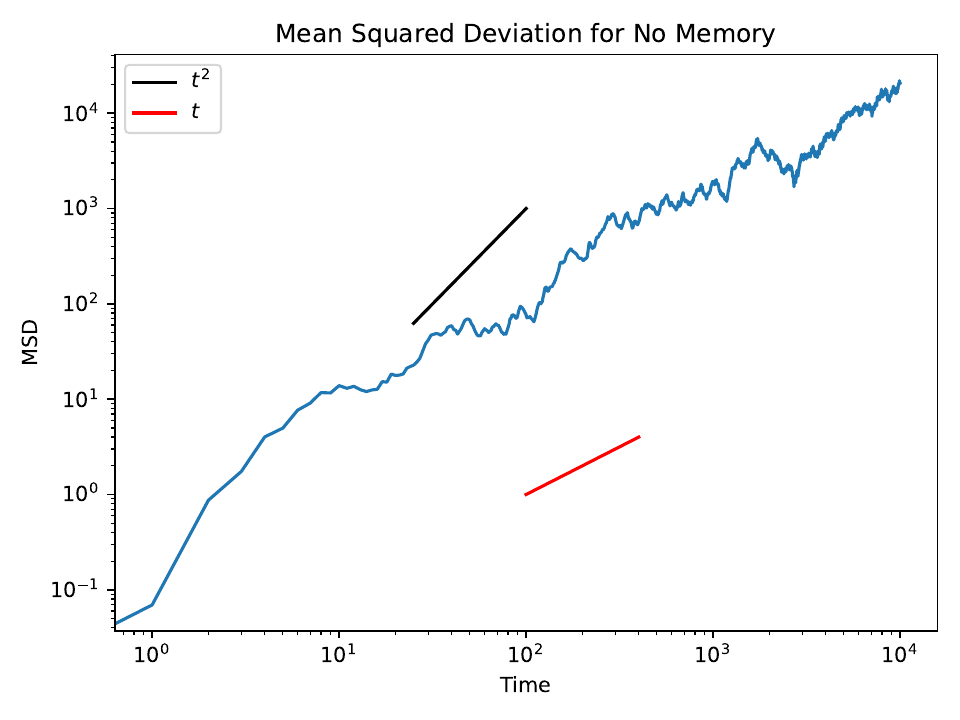}
		\caption{This figure shows the MSD calculated over 20 different paths for a version of Equation \ref{eq:fin-lang-eq} which has no memory.  The fractional order derivative goes to one, which returns Equation \ref{eq:no-memory}.}
		\label{fig:msd-nomem2}
\end{figure}

The third region, at the end, is the long-term behavior of the system.  In the case of $\alpha=0.2$, the dispersion goes as $t^{2-\alpha}$.  A discussion of the properties of this phase will follow.  Between the beginning exploratory behavior and the long-term behavior, there is a transition region.  Here, the MSD plateaus becoming a constant which implies more solid-like behavior.  We believe this is where the system is building up more and more memory.  The memory is still quite small, thus it has very little impact on the dynamics.  Once the memory has built up to a significant enough level, the system moves into the third region which displays the dynamics of a system with memory.  Figure \ref{fig:msd-nomem2} shows the MSD plot for a system with no memory and there is no middle transition region.  Therefore, the existence of a transition region appears tied to the inclusion of memory in the system.  For regular Brownian motion, the scaling relation with time is just linear and this is exactly the long-term behavior seen in Figure \ref{fig:msd-nomem2}.

From a financial perspective, this could mean that there is a reaction time and shape associated to a given financial system.  

\section{Long-term Phase Behavior}
\begin{figure}[h]
     \centering
     \begin{subfigure}[b]{0.48\textwidth}
         \centering
         \includegraphics[width=\textwidth]{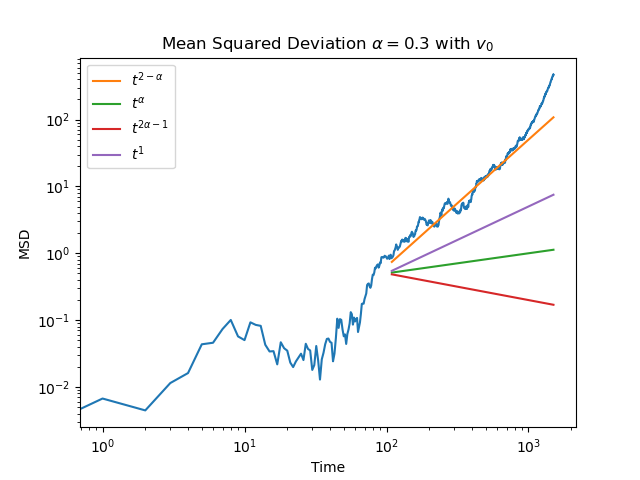}
         \subcaption{}
     \end{subfigure}
     \hfill
     \begin{subfigure}[b]{0.48\textwidth}
         \centering
         \includegraphics[width=\textwidth]{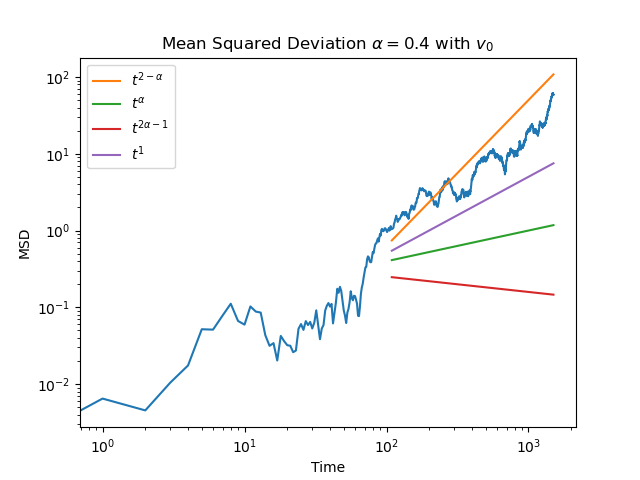}
         \subcaption{}
     \end{subfigure}
        \caption{This figure shows two MSD curves at two values of $\alpha$.  Around $\alpha=0.35$, there appears to be a change in the long-term dispersion behavior of the system.  The reference lines in both figures show the change from $t^{2-\alpha}$ to a scaling of $t^{\alpha}$.}
        \label{fig:msd-phases-trans}
\end{figure}

Figure \ref{fig:msd-phases-trans} shows the MSD curves for two values of $\alpha$ both with finite initial velocity.  In the third region, which corresponds to the long-term phase behavior, the two curves follow different dispersion relations.  For $\alpha=0.3$, it scales as $t^{2-\alpha}$ while for $\alpha=0.4$ it scales as $t^{\alpha}$.  It appears that around these values of $\alpha$ the dynamics of the dispersion change into a different form of anomalous diffusion.  In Figure \ref{fig:msd-phases-trans}, $\alpha=0.4$ does look close to linear scaling, but Table \ref{fig:all-msds} shows the full range of $\alpha$ parameters.  After $\alpha=0.4$, it can be seen that the MSD scales as $t^{\alpha}$ for the rest of the $\alpha$ values.

\subsection{Small $\alpha$ Behavior}
For small values of $\alpha$, as in Figure \ref{fig:msd-0p2}, there appears to to be kind of staircase shape to the long-term MSD.  Since the volatility can be calculated from the MSD, it appears that the volatility goes through repeating regions of flat volatility, then increasing volatility.  While the scaling of the long-term behavior shows anomalous diffusion, the overall staircase shape is similar to that of a marginal glass \cite{glass-phases}.  This region is not exactly a marginal glass, but some of its qualitative features are similar and so can be used to potentially infer what is physically taking place in the system.

The MSD appears to move from flatter regions, which are more glass-like behavior, to higher volatility regions which correspond to anomalous diffusion, and then back to glass-like again.  In a marginal glass, this behavior takes place because single particles cannot move on their own, but they become part of a group of particles that can move as a whole \cite{gardner}.  These cages have a fractal structure with cages inside of cages.  Because of this fractal structure, there exist metastable states for the particles given by the size of the cages.  When a region of anomalous diffusion reaches the next cage size, the MSD saturates and plateaus appear.  However, because the states are metastable, the plateau only exists for a finite period of time.  More study is required to see which parameters impact the lifetime of the plateaus and the cage sizes.  The existence of this qualitative behavior in the MSD shows that the topic is worth further investigation.  In order to properly characterize this phase, the asymptotic limits need to be understood, which would require more computational power.  A marginal glass is not stable in the long term because that would require infinite energy so these asymptotic limits are important \cite{robin-time-glass}.

It is difficult to say what these cages represent in a financial system as the system is very complex and we will not take a random guess with sufficient reasoning behind the analysis.  However, if these regions can be understood and recognized, it could lead to a better understanding of changes in market dynamics.

\begin{figure}[h]
     \centering
     \begin{subfigure}[b]{0.48\textwidth}
         \centering
         \includegraphics[width=\textwidth]{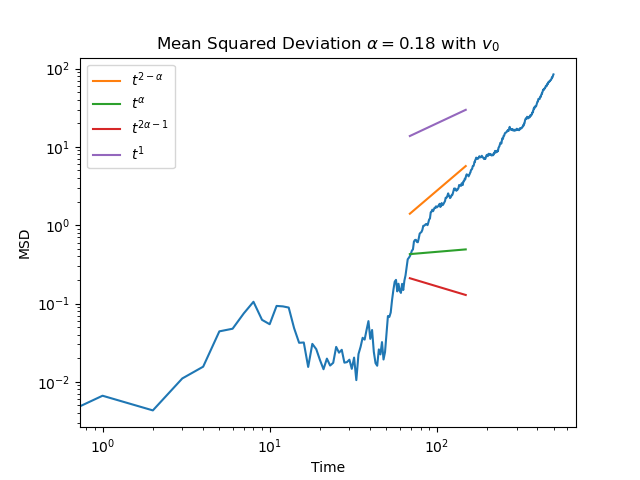}
         \subcaption{}
     \end{subfigure}
     \hfill
     \begin{subfigure}[b]{0.48\textwidth}
         \centering
         \includegraphics[width=\textwidth]{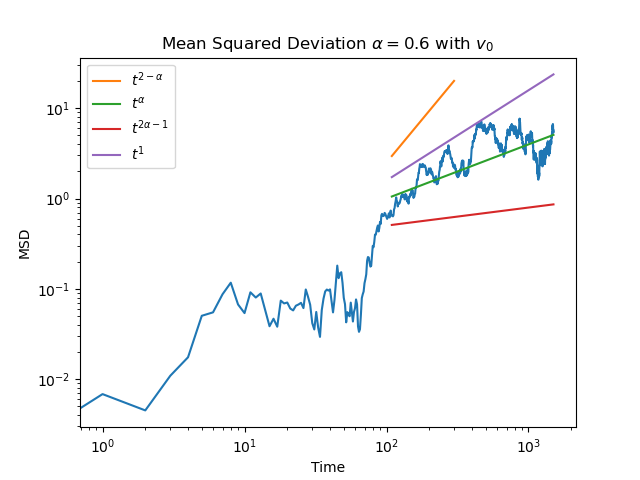}
         \subcaption{}
     \end{subfigure}
        \caption{The figures here show the MSD for two widely different values of the fractional derivative order, but with all other parameters of the differential equation kept the same.  It is clear that as $\alpha$ increases, there is more noise in the long-term MSD.}
        \label{fig:msd-oscils}
\end{figure}

\subsection{Oscillatory Behavior}
As the fractional derivative order increases, the strength of the memory increases as well; memory becomes more important.  Figure \ref{fig:msd-oscils} shows MSD plots for $\alpha=0.18$ and $0.6$ with all other parameters kept the same.  For larger values of $\alpha$, there appears to be larger oscillations in the MSD.  This leads us to believe that, in a rudimentary way, it is the memory that induces oscillations in the MSD.

There is not enough precision is our calculations to say whether these oscillations are truly periodic or if they are random.  However, the existence of large changes in the MSD can be seen as changes in the volatility over time.  Volatility is in part a reflection of the confidence users have in a given financial market; high volatility means users are unsure about the future while low volatility means high confidence in future expectations.  Because the volatility calculated in this way comes from historical information about the asset price, it is the realized volatility.

Another form of volatility is the implied volatility which can be calculated from options pricing models.  One of the most famous options pricing models is the Black-Scholes model which is based on a stochastic process that represents regular Brownian motion \cite{malliavin-book}.  Since the development of fractional Brownian motion, many options pricing models have been built using this method and are typically called rough volatility models.  These include the rough Heston model, the SABR model, the rough Bergomi model, along with many others \cite{rough-heston}-\cite{malliavin-book}.

The mathematical base of these models are stochastic differential equations, however the model described in this report has a different structure.  It is possible that our model is equivalent to one of the other stochastic models already developed, but it is a non-trivial task to show that equivalence.  By this same reasoning, it is also non-trivial to use the mathematical tools of other fractional volatility models to analyze the results found from our model.

\section{Asymptotic Behavior}
Given our expectations for the scaling of the MSD, we now analyze the simulation results more closely.  For $\alpha = 0.6$, the scaling appears to be closer to $t^{\alpha}$, however, as $\alpha$ grows the scaling starts to look more like $t^{2\alpha-1}$, for example $\alpha = 0.9$.  When $\alpha > 0.5$, it appears that the fractional derivative term is dominant where low values correspond to more interaction with the colored noise and higher values connect to the white noise.

For $\alpha = 0.2$ and $0.3$, the scaling is clearly $t^{2-\alpha}$, while as $\alpha$ increases the scaling gets closer to $t^{\alpha}$.  This leads us to believe that, qualitatively, the integer derivative interaction with colored noise leads to the $t^{2-\alpha}$ behavior.

The dispersion relation does transition between these different phases.  However the path by which it does this is unclear.  The complexity of having many terms in the differential equation means that it is difficult to analytically describe the transition behavior.  More extensive research would be required to analyze the equation and understand the expected asymptotic behavior and its transition as a function of $\alpha$.

\section{Conclusions}
The model reproduces the fundamental relationship between volatility and liquidity, as seen in Figure \ref{fig:volat-liquid}.  This relationship is unaffected by the existence of memory within the system.  For small values of the fractional derivative order, there exists behavior qualitatively similar to a marginal glass.  While this is far from conclusive, it shows some interesting features of financial markets that warrant further study.  Larger values of the fractional derivative order lead to oscillations in the realized volatility.  Because of the mathematical structure of the model it is difficult to use the typical tools of rough volatility financial models.  Therefore, more research would be required to compare this model with those already in existence to see if there are truly any different results.  The model analyzed in this report has generated potentially interesting behavior, but it has raised more questions than answers: continued study into this model is necessary to understand better what it is describing.

\newpage
\begin{table}[]
    \centering
    \begin{tabular}{cc}
        (a)\includegraphics[width=0.47\textwidth]{Report-Plots/fullcomp-msd-log-0p2.png} & (b)\includegraphics[width=0.47\textwidth]{Report-Plots/fullcomp-msd-log-0p3-long.png} \\
         (c)\includegraphics[width=0.47\textwidth]{Report-Plots/fullcomp-msd-log-0p4-long.png} & (d)\includegraphics[width=0.47\textwidth]{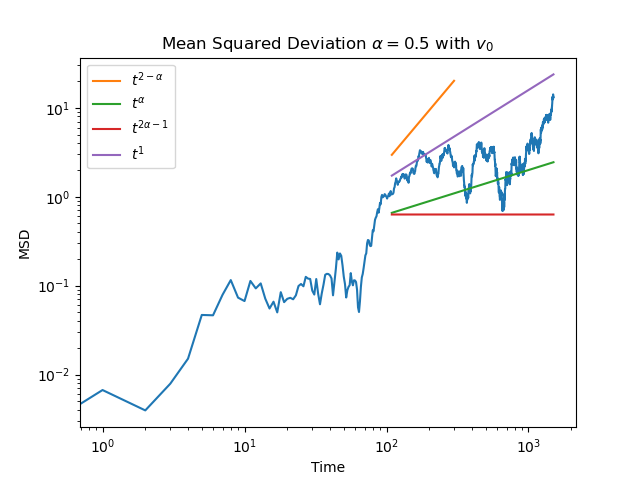} \\
         (e)\includegraphics[width=0.47\textwidth]{Report-Plots/fullcomp-msd-log-0p6-long.png} & (f)\includegraphics[width=0.47\textwidth]{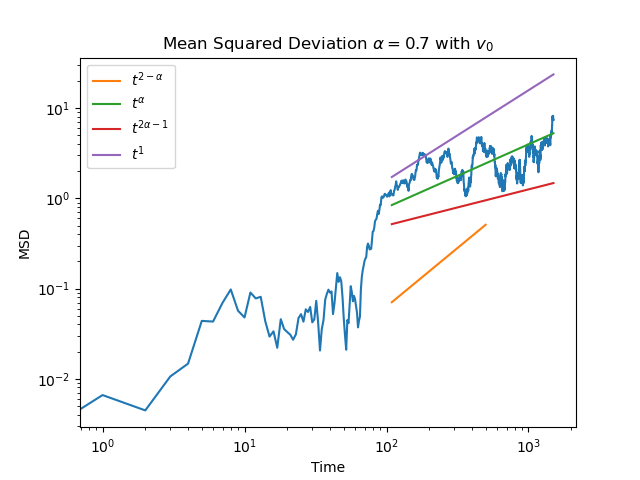} \\
         (g)\includegraphics[width=0.47\textwidth]{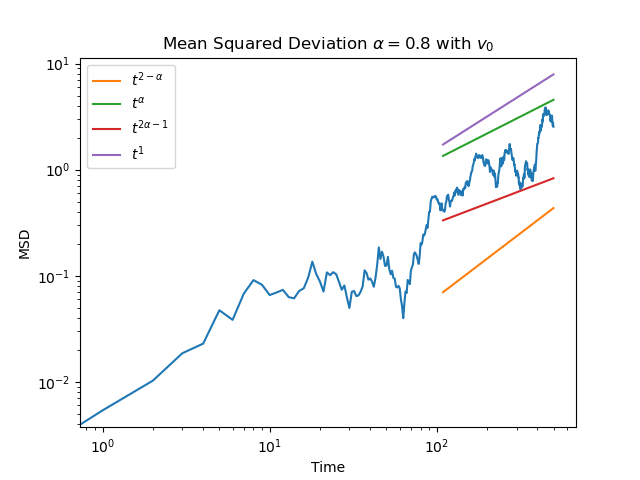} & (h)\includegraphics[width=0.47\textwidth]{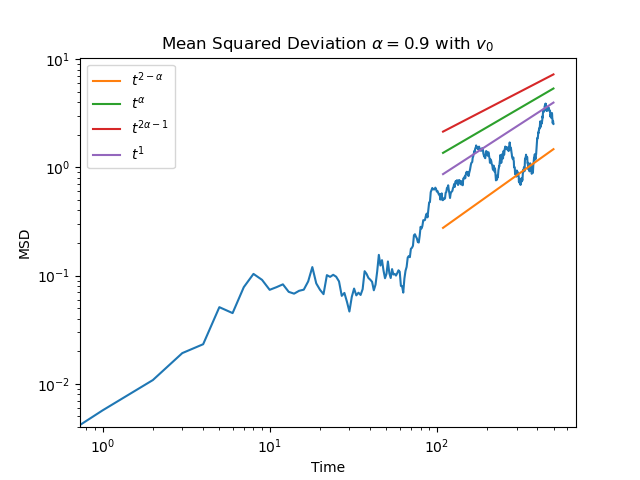}
    \end{tabular}
    \caption{In this table, the MSD for the full range of fractional derivative orders, $\alpha$, are made from solutions to Equation \ref{eq:fin-lang-eq} using the same input parameters: $\lambda=500$, $\beta=1.0=a$.}
    \label{fig:all-msds}
\end{table}

\chapter{Outlook}
In this research project, two different methods of generating fractional Brownian motion from physics and finance were brought together and shown to be equivalent.  Doing so required detailed analytical work showing the relationship between the fractional derivative, $\alpha$, from the differential equation-based physics approach and the Hurst parameter, $H$, from the stochastics-based finance approach: $\alpha=2-2H$.  Using numerics, fractional Brownian motion paths were generated by stochastic methods.  These paths were then differentiated and used as coloured noise in fractional differential equations.  In some cases, analytical solutions existed and where they were too difficult to find, a Monte Carlo method was developed from the ground up.

Starting with the financial differential equation given by Cont and Bouchard \cite{fin-lang}, a memory kernel was used that developed the equation into a fractional differential equation to describe a financial system with memory.  This memory-based financial system was then studied by finding solutions with the Monte Carlo method.  It was found that the system reproduced the expected relationship between market liquidity and volatility, and showed that the relationship is not impacted by the inclusion of memory in the model.

One helpful method for understanding systems of this type in physics is by examining the phases the system moves through.  By using this same method for analyzing a financial system, the system was seen to evolve through different phases, as memory is acquired by the system.  The dynamics of these phases are dependent on the fractional derivative order, $\alpha$, which corresponds to the strength of the correlation to the past.  Small values of $\alpha$ correspond to regions similar to that of anomalous Gardner phases, which oscillate in time between solid-like behavior and anomalous diffusion.  The data gathered about this phenomena is minimal and further study is required to understand what exactly is taking place in terms of phases.

With this physics-based analysis, the next logical step would be to see what these phases correspond to in terms of properties of financial systems.  However, the fractional differential equation used in this project is not written in the same mathematical language as most financial models, making it quite difficult to compare with others.  Further study would be needed to see if the model used here is fundamentally equivalent to other financial models or if it producing new dynamics.  The analysis here provides a base on which one can start a comparison work to see how a financial system with memory compares to those previously existing models.

While this research project has raised more questions than answers about financial models with memory, it does show a hint that these systems display intriguing properties that are worth further study.  The mathematical language of these two fields seems quite different but we have found that fundamentally they can be shown to be equivalent for Brownian motion.  The next step is to do so for more complex financial models with memory.

\end{document}